\documentclass[12pt, letterpaper]{JHEP3}

\usepackage{amssymb, amsmath, amsopn, amsthm}
\usepackage{latexsym}
\usepackage{graphics}
\usepackage{epsfig}

\title{Abelian and nonabelian vector field effective actions
 from string field theory}

\author{Erasmo Coletti, Ilya Sigalov, Washington Taylor \\
{Center for Theoretical Physics} \\ {MIT, Bldg.  6-308} \\
{Cambridge, MA 02139, U.S.A.} \\ {\rm email}: {\tt colettie,
sigalov, wati {\rm at domain} mit.edu}}

\abstract{The leading terms in the tree-level effective action for
the massless fields of the bosonic open string are calculated by
integrating out all massive fields in Witten's cubic string field
theory.  In both the abelian and nonabelian theories, field
redefinitions make it possible to express the effective action in
terms of the conventional field strength.  The resulting actions
reproduce the leading terms in the abelian and nonabelian
Born-Infeld theories, and include (covariant) derivative
corrections.}

\keywords{String Field Theory, Born-Infeld Action}
\preprint{MIT-CTP-3383, hep-th/0306041}

\newcommand{\junk}[1]{}

\newcommand{\NP}{{ Nucl.\ Phys.\ }}

\newcommand{\PL}{{Phys.\ Lett.\ }}
\newcommand{\PR}{{Phys.\ Rev.\ }}

\newcommand{\ket}[1]{\mathchoice{
    {\left|{#1}\right\rangle}}{|{#1}\rangle}{|{#1}\rangle}{|{#1}\rangle}}
\newcommand{\bra}[1]{\left\langle{#1}\right|}

\newcommand{\ld}{\lambda}
\newcommand{\pd}{\partial}

\newcommand{\half}{\frac{1}{2}}
\newcommand{\quarter}{\frac{1}{4}}

\newcommand{\cF}{\mathcal{F}}
\newcommand{\cH}{\mathcal{H}}
\newcommand{\cI}{\mathcal{I}}
\newcommand{\cJ}{\mathcal{J}}
\newcommand{\cM}{\mathcal{M}}
\newcommand{\cN}{\mathcal{N}}
\newcommand{\cO}{\mathcal{O}}

\newcommand{\cQ}{\mathcal{Q}}

\newcommand{\bZ}{\mathbb{Z}}
\newcommand{\bR}{\mathbb{R}}

\newcommand{\td}{\tilde}

\DeclareMathOperator{\Det}{Det} 
\DeclareMathOperator{\Log}{Log}

\DeclareMathOperator{\Tr}{Tr} 
\DeclareMathOperator{\tr}{tr}

\begin{document}
\section{Introduction}
Despite major advances in our understanding of nonperturbative
features of string theory and M-theory over the last eight years, we
still lack a fundamental nonperturbative and background-independent
definition of string theory.  String field theory seems to incorporate
some features of background independence which are missing in other
approaches to string theory.  Recent work, following the conjectures
of Sen \cite{Sen-universality}, has shown that Witten's open bosonic
string field theory successfully describes multiple distinct open
string vacua with dramatically different geometrical properties, in
terms of the degrees of freedom of a single theory (see
~\cite{Ohmori:2001am,DeSmet:2001af,Zwiebach:nj,Taylor:2002uv} for
reviews of this work).  An important feature of string field theory,
which allows it to transcend the usual limitations of local quantum
field theories, is its essential nonlocality.  String field theory is a
theory which can be defined with reference to a particular background
in terms of an infinite number of space-time fields, with highly
nonlocal interactions.  The nonlocality of string field theory is
similar in spirit to that of noncommutative field theories which have
been the subject of much recent work \cite{Douglas-Nekrasov}, but in
string field theory the nonlocality is much more extreme.  In order to
understand how string theory encodes a quantum theory of gravity at
short distance scales, where geometry becomes poorly defined, it is
clearly essential to achieve a better understanding of the nonlocal
features of string theory.

While string field theory involves an infinite number of
space-time fields, most of these fields have masses on the order
of the Planck scale.  By integrating out the massive fields, we
arrive at an effective action for a finite number of massless
fields.  In the case of a closed string field theory, performing
such an integration would give an effective action for the usual
multiplet of gravity/supergravity fields.  This action will,
however, have a complicated nonlocal structure which will appear
through an infinite family of higher-derivative terms in the
effective action.  In the case of the open string, integrating out
the massive fields leads to an action for the massless gauge
field.  Again, this action is highly nonlocal and contains an
infinite number of higher-derivative terms.  This nonlocal action
for the massless gauge field in the bosonic open string theory is
the subject of this paper.  By explicitly integrating out all
massive fields in Witten's open string field theory (including the
tachyon), we arrive at an effective action for the massless open
string vector field.  We compute this effective action
term-by-term using the level-truncation approximation in string
field theory, which gives us a very accurate approximation to each
term in the action.

It is natural to expect that the effective action we compute for the
massless vector field will take the form of the Born-Infeld action,
including higher-derivative terms.  Indeed, we show that this is the
case, although some care must be taken in making this connection.
Early work deriving the Born-Infeld action from string theory
\cite{Fradkin-Tseytlin85II,acny} used world-sheet methods \cite{
Fradkin-Tseytlin85-rus, Fradkin-Tseytlin85I}.  More recently, in the
context of the supersymmetric nonabelian gauge field action, other
approaches, such as $\kappa$-symmetry and the existence of
supersymmetric solutions, have been used to constrain the form of the
action (see \cite{k-Sevrin} for a recent discussion and further
references).  In this work we take a different approach.  We start with
string field theory, which is a manifestly off-shell formalism.  Our
resulting effective action is therefore also an off-shell action.
This action has a gauge invariance which agrees with the usual
Yang-Mills gauge invariance to leading order, but which has
higher-order corrections arising from the string field star product.  A
field redefinition analogous to the Seiberg-Witten map
\cite{Seiberg-Witten,Cornalba-Schiappa} is necessary to get a field
which transforms in the usual fashion \cite{Ghoshal-Sen,David-U}.  We
identify the leading terms in this transformation and show that after
performing the field redefinition our action indeed takes the
Born-Infeld form in the abelian theory.  In the nonabelian theory,
there is an additional subtlety, which was previously encountered in
related contexts in \cite{Ghoshal-Sen,David-U}.  Extra terms appear
in the form of the gauge transformation which cannot be removed by a
field redefinition.  These additional terms, however, are trivial and
can be dropped, after which the standard form of gauge invariance can
be restored by a field redefinition.
This leads to an effective action in the nonabelian theory which takes
the form of the nonabelian Born-Infeld action plus derivative
correction terms.

It may seem surprising that we integrate out the tachyon as well as
the fields in the theory with positive mass squared.  This is,
however, what is implicitly done in previous work such as
\cite{Fradkin-Tseytlin85II,acny} where the Born-Infeld action is
derived from bosonic string theory.  The abelian Born-Infeld action
can similarly be derived from recent proposals for the coupled
tachyon-vector field action \cite{Sen-action,Garousi,bddep,Kluson} by
solving the equation of motion for the tachyon at the top of the hill.
In the supersymmetric theory, of course, there is no tachyon on a BPS
brane, so the supersymmetric Born-Infeld action should be derivable
from a supersymmetric open string field theory by only integrating out
massive fields.  Physically, integrating out the tachyon corresponds
to considering fluctuations of the D-brane in stable directions, while
the tachyon stays balanced at the top of its potential hill.  While
open string loops may give rise to problems in the effective theory
\cite{Ellwood-Shelton-Taylor}, at the classical level the resulting
action is well-defined and provides us with an interesting model in
which to understand the nonlocality of the Born-Infeld action.  The
classical effective action we derive here must reproduce all on-shell
tree-level scattering amplitudes of massless vector fields in bosonic
open string theory.  To find a sensible action which includes quantum
corrections, it is probably necessary to consider the analogue of the
calculation in this paper in the supersymmetric theory, where there is
no closed string tachyon.

The structure of this paper is as follows: In Section 2 we review
the formalism of string field theory, set notation and make some
brief comments regarding the Born-Infeld action.  In Section 3 we
introduce the tools needed to calculate terms in the effective
action of the massless fields.  Section 4 contains a calculation
of the effective action for all terms in the Yang-Mills action.
Section 5 extends the analysis to include the next terms in the
Born-Infeld action in the abelian case and Section 6 does the same
for the nonabelian analogue of the Born-Infeld action.  Section 7
contains concluding remarks.  Some useful properties of
the Neumann matrices appearing in the 3-string vertex of Witten's
string field theory are included in the Appendix.

\section{Review of formalism}
Subsection \ref{sec:basics} summarizes our notation and the basics
of string field theory.  In subsection \ref{sec:amplitudesreview}
we review the method of \cite{Taylor-amplitudes} for computing
terms in the effective action.  The last subsection,
\ref{sec:Born-Infeld}, contains a brief discussion of the
Born-Infeld action.

\subsection{Basics of string field theory}
\label{sec:basics} In this subsection we review the basics of
Witten's open string field theory \cite{Witten-SFT}.  For further
background information see the
reviews~\cite{lpp,Gaberdiel-Zwiebach,Thorn:1988hm,Zwiebach:nj}.
The degrees of freedom of string field theory (SFT) are
functionals $\Phi[x (\sigma); c (\sigma), b (\sigma)]$ of the
string configuration $x^{\mu} (\sigma)$ and the ghost and
antighost fields $c (\sigma)$ and $b (\sigma) $ on the string at a
fixed time.  String functionals can be expressed in terms of
string Fock space states, just as functions in ${\cal L}^2 (\bR)$
can be expressed as linear combinations of harmonic oscillator
eigenstates.  The Fock module of a single string of momentum $p$
is obtained by the action of the matter, ghost and antighost
oscillators on the (ghost number one) highest weight vector
$\ket{p}$.  The action of the raising and lowering oscillators on
$\ket{p}$ is defined by the creation/annihilation conditions and
commutation relations
\begin{align}
a^{\mu}_{n\ge 1}\ket{p} & = 0,  &
[a^{\mu}_m, a^{\nu}_{-n}] & = \eta^{\mu \nu} \delta_{m,n}, \nonumber\\
p^{\mu}\ket{k}  = & k^{\mu}  \ket{k}, \\
b_{n\ge 0}\ket{p} & = 0,  &
\{b_m, c_{-n}\} & = \delta_{m,n}, \nonumber\\
c_{n\ge 1}\ket{p} & = 0.  \nonumber
\end{align}
Hermitian conjugation is defined by $a^{\mu\dagger}_n =
a^{\mu}_{-n}$, $b^\dagger_n = b_{-n}$, $c^\dagger_n = c_{-n}$.  The
single-string Fock space is then spanned by the set of all vectors
$\ket{\chi} = \cdots a_{n_2} a_{n_1} \cdots b_{k_2} b_{k_1}\cdots
c_{l_2} c_{l_1} \ket{p}$ with $n_i , k_i < 0 $ and $l_i \le 0$.
String fields of ghost number 1 can be expressed as linear
combinations of such states $\ket{\chi}$ with equal number of
$b$'s and $c$'s, integrated over momentum.
\begin{equation}
\label{eq:stringfields} \ket{\Phi}  = \int d^{26} p \; \left( \phi
(p) + A_\mu (p) \; a^\mu_{-1} - i \alpha(p) b_{-1} c_{0} +
B_{\mu\nu}(p) a_{-1}^\mu a_{-1}^\nu + \cdots \right)\ket{p}\,.
\end{equation}
The Fock space vacuum $\ket{0}$ that we use is related to the
$SL(2, \bR)$ invariant vacuum $\ket{1}$ by $\ket{0} = c_1
\ket{1}$.  Note that $\ket{0}$ is a Grassmann odd object, so that
we should change the sign of our expression whenever we
interchange $\ket{0}$ with a Grassmann odd variable.  The bilinear
inner product between the states in the Fock space is defined by
the commutation relations and
\begin{equation}
\bra{k} c_0\,   \ket{p}  = (2 \pi)^{26} \delta(k + p).
\end{equation}
The SFT action can be written as
\begin{equation}
S = -\frac{1}{2}\bra{V_2} \Phi, Q_B \Phi \rangle - \frac{g}{3}
\bra{V_3} \Phi, \Phi, \Phi \rangle \label{eq:action}
\end{equation}
where $\ket{V_n} \in {\cal H}^n$.  This action is invariant under
the gauge transformation
\begin{equation}
\delta\ket{\Phi}  = Q_B\ket{\Lambda} + g\bigl(\bra{\Phi, \Lambda}
V_{3}\rangle- \bra{\Lambda, \Phi} V_{3}\rangle\bigl)
\label{eq:gaugetransform}
\end{equation}
with $\Lambda$ a string field gauge parameter at ghost number 0.
Explicit oscillator representations of $\bra{V_2}$ and $\bra{V_3}$
are given by \cite{Gross-Jevicki12, cst, Samuel, Ohta}
\begin{equation}
\bra{V_2}  =   \int d^{26}  p \; \bra{p}^{\text{(1)}}
\otimes{\bra{-p}}^{\text{(2)}} \bigl ( c_0^{(1)}+ c_0^{(2)} \bigr)
\exp\left(a^{(1)}\cdot C \cdot a^{(2)} - b^{(1)} \cdot C \cdot
c^{(2)} - b^{(1)} \cdot C \cdot c^{(2)} \right) \label{eq:v2}
\end{equation}
and
\begin{multline}
\bra{V_3}  = \cN \int \prod_{i = 1}^3 \left(d^{26} p_i {\bra{p}}^{(i)}
c_0^{(i)} \right )
\delta \bigl (\sum p_j\bigr) \\
\times \exp\left( \half a^{(r)} \cdot V^{r s} \cdot a^{(s)} -
p^{(r)} V^{rs}_{0 \cdot} \cdot a^{(s)} + \half p^{(r)} V^{rr}_{00}
p^{(r)} - b^{(r)}\cdot X^{r s} \cdot c^{(s)}\right ) \label{eq:v3}
\end{multline}
where all inner products denoted by $\cdot$ indicate summation
from 1 to $\infty$ except in $b \cdot X$, where the summation
includes the index 0.  The contracted Lorentz indices in $a_n^\mu$
and $p_\mu$ are omitted.  $C_{mn}=(-1)^n\delta_{mn}$ is the BPZ
conjugation matrix.  The matrix elements $V^{rs}_{mn}$ and
$X^{rs}_{mn}$ are called Neumann coefficients.  Explicit
expressions for the Neumann coefficients and some relevant
properties of these coefficients are summarized in the Appendix.
The normalization constant $\cN$ is defined by 
\begin{equation}
\cN = \exp(- \half
\sum_r V_{00}^{rr})= \frac{3^{9/2}}{2^6} ,
\end{equation}
so that the on-shell three-tachyon amplitude
is given by $2 g$.  We use units where $\alpha' = 1$.

\subsection{Calculation of effective action}
\label{sec:amplitudesreview} String field theory can be thought of
as a (nonlocal) field theory of the infinite number of fields that
appear as coefficients in the oscillator expansion
(\ref{eq:stringfields}).  In this paper, we are interested in
integrating out all massive fields at tree level.  This can be
done using standard perturbative field theory methods.  Recently
an efficient method of performing sums over intermediate particles
in Feynman graphs was proposed in \cite{Taylor-amplitudes}.  We
briefly review this approach here; an alternative approach to such
computations has been studied recently in \cite{Bars}.

In this paper, while we include the massless auxiliary field
$\alpha$ appearing in the expansion  (\ref{eq:stringfields}) as
an external state in Feynman diagrams, all the massive fields we
integrate out are contained in the Feynman-Siegel gauge string
field satisfying
\begin{equation}
\label{eq:Feynman-Siegel} b_0\ket{\Phi} = 0,
\end{equation}
This means that intermediate states in the tree diagrams we
consider do not have a $c_0$ in their oscillator expansion.  For
such states, the propagator can be written in terms of a Schwinger
parameter $\tau$ as
\begin{equation}
\frac{b_0}{L_0} = b_0 \int_0^\infty d \tau \; e^{-\tau L_0},
\label{eq:propagator}
\end{equation}
In string field theory, the Schwinger parameters can be
interpreted as moduli for the Riemann surface associated with a
given diagram
\cite{Giddings:1986bp,Giddings:1986wp, Thorn-perttheory, Thorn:1988hm,
Zwiebach:1990az}.

In field theory one computes amplitudes by contracting vertices
with external states and propagators.  Using the quadratic and
cubic vertices (\ref{eq:v2}), (\ref{eq:v3}) and the propagator
(\ref{eq:propagator}) we can do same in string field theory.  To
write down the contribution to the effective action arising from a
particular Feynman graph we include a vertex $\bra{V_3} \in
\cH^{*3}$ for each vertex of the graph and a vertex $\ket{V_2}$
for each internal edge.  The propagator (\ref{eq:propagator}) can
be incorporated into the quadratic vertex through
\footnote{Consider the tachyon propagator as an example.  We
contract $c_0 \ket{p_{1}}$ and $c_0 \ket{p_{2}}$ with  $\bra{P}$
to  get
\begin{equation}
\label{eq:tachprop}
 \bra{P} c_0\ket{p_{1}} c_0\ket{p_{2}} =
- \int_0^\infty d \tau e^{\tau (1 - p_{1}^2)}\delta(p_{1}+p_{2})=
-\frac{\delta(p_{1}+p_{2})} {p_{1}^2-1}.
\end{equation}
This formula assumes that both momenta are incoming.  Setting $p_1
= - p_2 = p$ and using the metric with  $(-,+,+,...,+)$ signature
we have
\begin{equation}
-\frac{1}{p^2+m^2}=\frac{1}{p_{0}^2 - \vec{p}^2 - m^2}
\end{equation}
thus (\ref{eq:tachprop}) is indeed the correct propagator for the
scalar particle of mass $m^2 = -1$.}
\begin{equation}
\ket{P} = - \int_0^\infty d \tau \; e^{\tau (1 - p^2)} \ket{\td
V_2}.
\end{equation}
where in the modified vertex  $\ket{\tilde{V}_2 (\tau)}$ the ghost
zero modes $c_0$ are canceled by the $b_0$ in
(\ref{eq:propagator}) and the matrix $C_{mn}$ is replaced by
\begin{equation}
\tilde{C}_{mn} (\tau) = e^{-m \tau} (-1)^m \delta_{mn} \,.
\end{equation}
With these conventions, any term in the effective action can be
computed by contracting the three-vertices from the corresponding
Feynman diagram on the left with factors of $| P \rangle$ and
low-energy fields on the right (or vice-versa, with $| V_3
\rangle$'s on the right and $\langle P |$'s on the left).  Because the
resulting expression integrates out {\it all}\, Feynman-Siegel gauge
fields along interior edges, we must remove the contribution from
the intermediate massless vector field by hand when we are
computing the effective action for the massless fields.  Note that
in \cite{Taylor-amplitudes}, a slightly different method was used
from that just described; there the propagator was incorporated
into the three-vertex rather than the two-vertex.  Both methods
are equivalent; we use the method just described for convenience.

States of the form
\begin{equation}
\exp\left(\ld \cdot a^\dagger  + \half a^\dagger \cdot S \cdot
a^\dagger \right) \ket{p}
\end{equation}
are called squeezed states.  The vertex $\ket{V_3}$ and the
propagator $\ket{P}$ are (linear combinations of) squeezed states
and thus are readily amenable to computations.  The inner product
of two squeezed states is given by \cite{Kostelecky-Potting}
\begin{multline}
\bra{0}\exp(\ld\cdot a + \half a\cdot S\cdot a) \exp(\mu\cdot
a^{\dagger} + \half a^{\dagger}\cdot V\cdot
a^{\dagger})\ket{0}  \\
= \Det(1-S\cdot V)^{-1/2} \exp  \bigl[\ld\cdot(1-V\cdot S)^{-1} \cdot \mu \\
+ \half \ld\cdot(1-V\cdot S)^{-1}\cdot V \cdot \ld + \half \mu
\cdot S \cdot(1-V\cdot S)^{-1}\cdot \mu\bigr]
\label{eq:mattersqueezed}
\end{multline}
and (neglecting ghost zero-modes)
\begin{multline}
\bra{0}\exp(b\cdot\ld_b - \ld_c\cdot c  - b\cdot S\cdot c)
\exp(b^\dagger\cdot\mu_b + \mu_c\cdot c^\dagger  +  b^\dagger\cdot
V\cdot c^\dagger)
\ket{0} \\
= \Det(1-S\cdot V) \exp\bigl[- \ld_c \cdot(1-V\cdot S)^{-1}\cdot
\mu_b -
 \mu_c  \cdot(1 -S\cdot V)^{-1}\cdot  \ld_b \\
+ \ld_c\cdot(1-V\cdot S)^{-1}\cdot V \cdot \ld_b + \mu_c\cdot S
\cdot(1-V\cdot S)^{-1}\cdot \mu_b\bigr].  \label{eq:ghostsqueezed}
\end{multline}
Using these expressions, the combination of three-vertices and
propagators associated with any Feynman diagram can be simply
rewritten as an integral over modular (Schwinger) parameters of a
closed form expression in terms of the infinite matrices $V_{nm},
X_{nm}, \tilde{C}_{nm} (\tau)$.  The schematic form of these
integrals is
\begin{multline}
(\bra{V_3})^v (\ket{P})^i \sim \left(\prod_{j = 1}^{i} \int \; d
\tau^j \right)
\frac{\Det (1-\hat{C} \hat{X})}{ \Det (1-\hat{C} \hat{V})^{13}} \\
\times
(\langle 0 |)^{3v-2i} \exp\left( \half a^\dagger\cdot S\cdot a^\dagger + \mu
\cdot a^{\dagger}
 + b^\dagger \cdot U \cdot  c^\dagger+ \mu_c \cdot c^{\dagger}
+ b^{\dagger} \cdot \mu_b \right) 
\end{multline}
where $\hat{C}, \hat{X}, \hat{V}$ are matrices with blocks of the
form $\tilde{C}, X, V$ arranged according to the combinatorial
structure of the diagram.  The matrix $\hat{C}$ and the squeezed
state coefficients $S, U, \mu, \mu_b, \mu_c$ depend implicitly on
the modular parameters $\tau^i$.

\subsection{The effective vector field action and Born-Infeld}
\label{sec:Born-Infeld}

In this subsection we describe how the effective
action for the vector field is determined from SFT, and we discuss
the Born-Infeld action \cite{Born-Infeld} which describes the leading 
terms in this effective action.  For a more detailed review of the Born-Infeld
action, see \cite{Tseytlin}

As discussed in subsection 2.1, the string field theory action is a
space-time action for an infinite set of fields, including the
massless fields $A_\mu (x)$ and $\alpha (x)$.  This action has a very
large gauge symmetry, given by (\ref{eq:gaugetransform}).  We wish to
compute an effective action for $A_\mu (x)$ which has a single gauge
invariance, corresponding at leading order to the usual Yang-Mills
gauge invariance.  We compute this effective action in several
steps.  First, we use Feynman-Siegel gauge (\ref{eq:Feynman-Siegel})
for all {\it massive} fields in the theory.  This leaves a single
gauge invariance, under which $A_\mu$ and $\alpha$ have linear
components in their gauge transformation rules.  This partial gauge
fixing is described more precisely in section
\ref{sec:fieldredefinitions}.  Following this partial gauge fixing, all
massive fields in the theory, including the tachyon, can be integrated
out using the method described in the previous subsection, giving an
effective action
\begin{equation}
\check{S} [A_\mu (x), \alpha (x)] 
\end{equation}
depending on $A_\mu$ and $\alpha$.  We can then further integrate out
the field $\alpha$, which has no kinetic term, to derive the desired
effective action
\begin{equation}
S[A_\mu (x)] \,.  \label{eq:s}
\end{equation}
The action (\ref{eq:s}) still has a gauge invariance, which at
leading order agrees with the Yang-Mills gauge invariance
\begin{equation}
\delta A_\mu (x) = \partial_\mu \lambda (x) -ig_{{\rm YM}}[A_\mu
(x),
  \lambda (x)] +
\cdots \label{eq:transform}
\end{equation}

The problem of computing the effective action for the massless gauge
field in open string theory is an old problem, and has been addressed
in many other ways in past literature.  Most methods used in the past
for calculating the effective vector field action have used
world-sheet methods.  While the string field theory approach we use
here has the advantage that it is a completely off-shell formalism, as
just discussed the resulting action has a nonstandard gauge invariance
\cite{David-U}.  In world-sheet approaches to this computation, the
vector field has the standard gauge transformation rule
(\ref{eq:transform}) with no further corrections.  A general theorem
\cite{Henneaux-db} states that there are no deformations of the
Yang-Mills gauge invariance which cannot be taken to the usual
Yang-Mills gauge invariance by a field redefinition.  In accord with
this theorem, we identify in this paper field redefinitions which take
the massless vector field $A_\mu$ in the SFT effective action
(\ref{eq:s}) to a gauge field $ \hat{A}_\mu$ with the usual gauge
invariance.  We write the resulting action as
\begin{equation}
\hat{S}[\hat{A}_\mu (x)] \,.  \label{eq:sh}
\end{equation}
This action, written in terms of a conventional gauge field, can
be compared to previous results on the effective action for the
open string massless vector field.

Because the mass-shell condition for the vector field $A_\mu (p)$
in Fourier space is $p^2 = 0$, we can perform a sensible expansion
of the action (\ref{eq:s}) as a double expansion in $p$ and $A$.
We write this expansion as
\begin{equation}
S[A_\mu] = \sum_{n = 2}^{ \infty}  \sum_{k = 0}^{ \infty}
S^{[k]}_{A^n}  \label{eq:expansion}
\end{equation}
where $S^{[k]}_{A^n}$ contains the contribution from all terms of
the form $\partial^kA^n$.  A similar expansion can be done for
$\hat{S}$, and we similarly denote by
$\check{S}^{[k]}_{\alpha^mA^n}$ the sum of the terms in
$\check{S}$ of the form $\partial^k \alpha^mA^n$.

Because the action $\hat{S}[\hat{A}]$ is a function of a gauge
field with conventional gauge transformation rules, this action
can be written in a gauge invariant fashion; {\it i.e.} in terms
of the gauge covariant derivative $\hat{D}_\mu = \partial_\mu
-ig_{{\rm YM}}[\hat{A}, \cdot]$ and the field
strength $\hat{F}_{\mu \nu}$.  For the abelian theory,
$\hat{D}_\mu$ is just $\partial_\mu$, and there is a natural
double expansion of $\hat{S}$ in terms of $p$ and $F$.  It was
shown in \cite{Fradkin-Tseytlin85II,acny} that in the abelian
theory the set of terms in $\hat{S}$ which depend only on
$\hat{F}$, with no additional factors of $p$ ({\it i.e.}, the
terms in $\hat{S}^{[n]}_{\hat{A}^n} $) take the Born-Infeld form
(dropping hats)
\begin{equation}
S_{BI}=-\frac{1}{(2\pi g_{YM})^{2}} \int d x
\sqrt{-\det\left(\eta_{\mu\nu} + 2\pi g_{YM} F_{\mu\nu}\right)}
\label{eq:BIaction}
\end{equation}
where
\begin{equation}
F_{\mu\nu}=\pd_{\mu}A_{\nu}-\pd_{\nu}A_{\mu}
\end{equation}
is the gauge-invariant field strength.  Using $\log \left(\det
M\right) = \tr\left(\log(M)\right)$ we can expand in $F$ to get
\begin{multline}
S_{BI} = -\frac{1}{(2 \pi g_{YM})^{2}} \int d x  \biggl(1 +
\frac{(2\pi g_{YM})^2}{4} F_{\mu\nu}F^{\mu\nu} \\ - \frac{(2\pi
g_{YM})^4}{8} \left ( F_{\mu \nu} F^\nu_{\ \ld} F^\ld_{\ \sigma}
F^{\sigma\mu} - \quarter (F_{\mu \nu} F^{\mu\nu})^2 \right ) +
\cdots \bigg).  \label{eq:BIexpansion}
\end{multline}
We expect that after the appropriate field redefinition, the
result we calculate from string field theory for the effective
vector field action (\ref{eq:s}) should contain as a leading part
at each power of $\hat{A}$ terms of the form
(\ref{eq:BIexpansion}), as well as higher-derivative terms of the
form $\partial^{n + k}A^n$ with $k > 0$.  We show in section 5
that this is indeed the case.

The nonabelian theory is  more complicated.  In the nonabelian
theory we must include covariant derivatives, whose commutators
mix with field strengths through relations such as
\begin{equation}
[D_\mu, D_\nu] F_{\ld \sigma}= [F_{\mu \nu}, F_{\ld \sigma}]   \,.
\end{equation}
In this case, there is no systematic double expansion in powers of $D$
and $F$.  It was pointed out by Tseytlin in \cite{Tseytlin97} that
when $F$ is taken to be constant, and both commutators $[F, F]$ and
covariant derivatives of field strengths $DF$ are taken to be
negligible, the nonabelian structure of the theory is irrelevant.  In
this case, the action reduces to the Born-Infeld form
(\ref{eq:BIaction}), where the ordering ambiguity arising from the
matrix nature of the field strength $F$ is resolved by the symmetrized
trace (STr) prescription whereby all possible orderings of
the $F$'s are averaged over.  While this observation is correct, it
seems that the symmetrized trace formulation of the nonabelian
Born-Infeld action misses much of the important physics of the full
vector field effective action.  In particular, this simplification of
the action gives the wrong spectrum around certain background fields,
including those which are T-dual to simple intersecting brane
configurations \cite{Hashimoto-Taylor,Bain,dst,bsttv}.  It seems that
the only systematic way to deal with the nonabelian vector field
action is to include all terms of order $F^n$ at once, counting $D$ at
order $F^{1/2}$.  The first few terms in the nonabelian vector field
action for the bosonic theory were computed in
\cite{Neveu-Scherk,
Scherk-Schwarz, Tseytlin86}.
The terms in the action up to $F^4$ are given
by
\begin{equation}
\label{eq:BI+DC} S_{\rm nonabelian} = \int -\quarter {\rm Tr}\;
F^2 + \frac{2 i g_{{\rm YM}}}{3}\Tr \left( F^3 \right) + \frac{(2
\pi g_{YM})^2} {8}{\rm STr}\;
 \left(F^4 - \quarter (F^2)^2 \right)
+ \cdots
\end{equation}
In section 6, we show that the effective action we derive from string
field theory agrees with (\ref{eq:BI+DC}) up to order $F^3$ after the
appropriate field redefinition .

\section{Computing the effective action}
In this section we develop some tools for calculating low-order
terms in the effective action for the massless fields by
integrating out all massive fields.  Section
\ref{sec:genfunctions} describes a general approach to computing
the generating functions for terms in the effective action and
gives explicit expressions for the generating functions of cubic
and quartic terms.  Section \ref{sec:masslessamps} contains a
general derivation of the quartic terms in the effective action
for the massless fields.  Section \ref{sec:truncation} describes
the  method we use to numerically approximate the coefficients in
the action.
\subsection{Generating functions for terms in the effective action}
\label{sec:genfunctions} A convenient way of calculating SFT
diagrams is to  first compute the off-shell amplitude with generic
external coherent states
\begin{equation}
\label{eq:cohstate} \ket{G} = \exp \left( J_{m \mu} a^{\mu}_{-m}
-b_{-m} \cJ_{b m}+ \cJ_{c m} c_{-m} \right) \ket{p}
\end{equation}
where the index $m$  runs from $1$ to $\infty$ in $J_m^\mu$ and $\cJ_{bm}$ 
and from $0$ to $\infty$ in $\cJ_{cm}$.

Let $\Omega_M({p^i, J^i, \cJ_b^i, \cJ_c^i;\, 1 \le i \le M})$ be
the sum of all connected tree-level diagrams with $M$ external
states $\ket{G^i}$.  $\Omega_M$ is a generating function for all
tree-level off-shell $M$-point amplitudes and can be used to
calculate all terms we are interested in in the effective action.
Suppose that we are interested in a term in the effective action
whose  $j$'th field  $\psi^{(j)}_{\mu_1, \ldots, \mu_{N}}(p)$ is
associated with the Fock space state
\begin{equation}
\label{eq:extstate} \prod_{m,n,q} a^{\mu_m}_{i_m} b_{k_n} c_{l_q}
\ket{p}.
\end{equation}
We can obtain the associated off-shell amplitude by acting on
$\Omega_M$ with the corresponding differential operator for each
$j$
\begin{equation}
\int d p \,  \psi^{(j)}_{\mu_1, \ldots, \mu_N}(p) \prod_{m,n,q}
\frac{\pd}{\pd J^{j}_{i_m \mu_m}} \frac{\pd}{\pd \cJ_{b k_n}^{j}}
\frac{\pd}{\pd \cJ_{c l_q}^{j}}
\end{equation}
and setting $J^j$, $\cJ^j_b$, and $\cJ^j_c$ to $0$.  Thus, all the
terms in the effective action which we are interested in can be
obtained from $\Omega_M$.

When we calculate a certain diagram with external states $| G^i
\rangle$ by applying formulae (\ref{eq:mattersqueezed}) and
(\ref{eq:ghostsqueezed}) for inner products of coherent and
squeezed states the result has the general form
\begin{multline}
\label{eq:generalanswer} \Omega_M =\delta\bigl(\sum p^r\bigr) \int
\prod_{\ell=1}^{N_{\text{prop}}} d\tau_\ell\,
 \cF(p, \tau) \\
\times \exp\left(\half J^i_{m}\Delta^{ij}_{mn}(\tau) J^j_{n} - p^i
\Delta^{ij}_{0m}(\tau) J^j_m + p_i \Delta^{ij}_{00}(\tau) p_j +
{\rm  ghosts} \right).
\end{multline}
A remarkable feature is that (\ref{eq:generalanswer}) depends on
the sources $J^{j}, \cJ^{j}_b, \cJ^{j}_c$ only through the
exponent of a quadratic form.  Wick's theorem is helpful in writing
the derivatives of the exponential in an efficient way.  Indeed,
the theorem basically reads
\begin{equation}
\prod_{i=1}^M \frac{\pd}{\pd J_{n}^i} \exp\left(\half J_m^j
\Delta_{mn}^{jk} J_n^k  \right) \Bigg |_{J^i_n =0} =\ \text{Sum over all
contraction products}
\end{equation}
where the sum is taken over all pairwise contractions, with the
contraction between $(n, i)$ and $(m, j)$ carrying the factor
$\Delta_{nm}^{ij}$.

Note that $\Omega_M$ includes contributions from all the
intermediate fields in Feynman-Siegel gauge.  To compute the
effective action for $A_\mu$ we must project out the contribution
from intermediate $A_\mu$'s.

\subsubsection{Three-point generating function}
\label{sec:3pointgenfunc} Here we illustrate the idea sketched
above with the simple example of the three-point generating
function.  This generating function provides us  with an efficient
method of computing the coefficients of the SFT action and the SFT
gauge transformation.  Plugging  $\ket{G^i}$ , $1 \le i \le 3$
into the cubic vertex (\ref{eq:v3}) and using
(\ref{eq:mattersqueezed}), (\ref{eq:ghostsqueezed}) to evaluate
the inner products we find
\begin{equation}
\label{eq:Omega3} \Omega_3 = - \frac{\cN g}{3} \,
\delta\bigl(\sum_r p^r\bigr) \exp\Bigl( \half p^{r} V^{rs}_{00}
p^{s} - p^{r} V^{rs}_{0n} J^{s}_n + \half J^{r}_m V^{rs}_{mn}
J^{s}_n -\cJ_{cm}^r X^{rs}_{mn} \cJ_{bn}^s \Bigr).
\end{equation}
As an illustration of how this generating function can be used
consider the three-tachyon term in the effective action.  The
external tachyon state is $\int d p \, \phi(p) \ket{p}$.  The
three-tachyon vertex is obtained from (\ref{eq:Omega3}) by simple
integration over momenta and setting the sources to 0.  No
differentiations are necessary in this case.  The three-tachyon
term in the action is then
\begin{equation}
- \frac{g}{3}\bra{V_3} \phi, \phi, \phi \rangle = - \frac{\cN
g}{3}\int  \delta(\sum_{s} p^{s}) \prod_{r} d p^{r} \phi(p_r)
\exp\left(  \half p^{r} V^{rs}_{00} p^{s} \right) = - \frac{\cN
g}{3} \int d x \, \td \phi(x)^3 \label{eq:3tach}
\end{equation}
where
\begin{equation}
\td \phi(x) = \exp\left(-\half V^{11}_{00} \pd^2 \right) \phi(x).
\label{eq:tilde}
\end{equation}
For on-shell tachyons, $\pd^2\phi (x) = - \phi(x)$, so that we
have
\begin{equation}
- \frac{g}{3}\bra{V_3} \phi, \phi, \phi \rangle = - \frac{g}{3}
\cN e^{\frac{3}{2} V^{11}_{00}} \int d x \phi(x)^3 = - \frac{g}{3}
\int d x \phi(x)^3.
\end{equation}
The normalization constant cancels so that the on-shell
three-tachyon amplitude is just $2 g$, in agreement with
conventions used here and in \cite{Polchinski}.

\subsubsection{Four-point generating function}
\label{sec:4pointgenfunc} 
\begin{figure}
\centering
\begin{picture}
(200,100)(- 100,- 50) \put(-40,0){\line(1,0){80}}
\put(-40,0){\line( -1, -1){30}} \put(-40,0){\line( -1, 1){30}}
\put(40,0){\line(1, 1){30}} \put(40,0){\line(1, -1){30}} \put(0,
0){\qbezier(-110, -15)(-113,0)(-110, 15)} \put(0, 0){\qbezier(110,
-15)(113,0)(110, 15)} \put(0, 0){\qbezier(-50, 45)(0, 50)(50, 45)}
\put(-110, 15){\vector(1, 4){3}} \put(-110, -15){\vector(1,
-4){3}} \put(110, 15){\vector(-1, 4){3}} 
\put(110, -15){\vector(-1, -4){3}} 
\put(-50, 45){\vector(-4, -1){3}} 
\put(50, 45){\vector(4, -1){3}} 
\put(-120, 0){\makebox(0,0){$T'$}}
\put(120, 0){\makebox(0,0){$T$}} \put(0, 55){\makebox(0,0){$R$}}
\put(80,30){\makebox(0,0){$\ \ \ket{G^1}$}}
\put(80,-30){\makebox(0,0){$\ \ \ket{G^2}$}}
\put(-80,30){\makebox(0,0){$\bra{G^3}\ \ $}}
\put(-80,-30){\makebox(0,0){$\bra{G^4}\ \ $}}
\end{picture}
\caption[x]{\footnotesize Twists $T$, $T'$ and reflection $R$ are
symmetries of the amplitude.} \label{f:quartic_diagrams}
\end{figure}
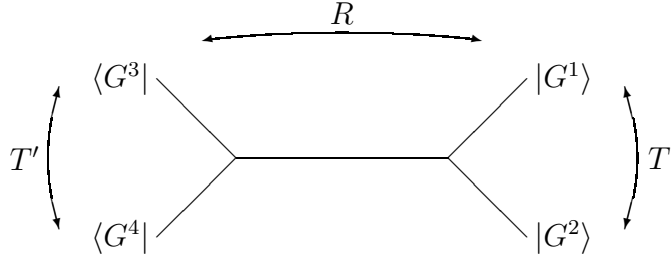

Now let us consider the generating
function for all quartic off-shell amplitudes (see
Figure~\ref{f:quartic_diagrams}).  The amplitude $\Omega_4$ after
contracting all indices can be written as
\begin{equation}
\label{eq:o4} \Omega_4 = \frac{\cN^2 g^2}{2}\, \int_0^\infty d\tau \;
e^{\tau(1-(p_1+p_2)^2)} \bra{\td V_2}\ket{R(1, 2)} \ket{R(3, 4)}
\end{equation}
where
\begin{equation}
\label{eq:rstate} \ket{R(i, j)}^{(k)} = \bra{G^i} \bra{G^j}
\ket{V_3}^{(ijk)}.
\end{equation}
Applying (\ref{eq:mattersqueezed}), (\ref{eq:ghostsqueezed}) to
the inner products in (\ref{eq:rstate}) we get
\begin{multline}
\ket{R(1, 2)} = \exp\bigl (\half p^\alpha_\mu U^{\alpha\beta}_{00}
p^{\mu \beta} - p^\alpha_\mu U^{\alpha \beta}_{0n} J^{\alpha
\mu}_n + \half J^\alpha_{m \mu} U^{\alpha \beta}_{mn} J^{\beta
\mu}_n \\ + a^{(3)}_{-m \mu} U^{33}_{mn} a^{\mu (3) }_{-n} +
(J^\alpha_{m \mu} U^{\alpha 3}_{mn} - p^\alpha_\mu
U^{\alpha3}_{0n}) a^{\mu (3)}_{-n} - \cJ_{cm}^\alpha
X^{\alpha\beta}_{mn} \cJ_{bn}^\beta \\ + b^{(3)}_{-m}
X^{3\alpha}_{mn} \cJ^\alpha_{bn} - \cJ_{cm}^\alpha
X^{\alpha3}_{mn} c^{(3)}_{-n} - b^{(3)}_{-m} X^{33}_{mn}  c^{(3)}_{-n}
 \bigr)c_0 \ket{-p^1-p^2}.
\end{multline}
Here  $\alpha, \beta \in {1, 2}$ and
\begin{equation}
U^{r s} =
\begin{pmatrix}
\label{eq:udef} V^{rs}_{0 0} - V^{r3}_{00} - V^{3 s}_{00} +
V^{33}_{00} \ \   &
V^{rs}_{0 n} - V^{3 s}_{0n}                                   \\
V^{rs}_{n 0} - V^{r 3}_{n0}  &  V^{rs}_{m n}
\end{pmatrix}.
\end{equation}
Using  (\ref{eq:mattersqueezed}), (\ref{eq:ghostsqueezed}) one
more time to evaluate the inner products in (\ref{eq:o4}) we
obtain
\begin{multline}
\label{eq:4genfuncprelim} \Omega_4 = \frac{\cN^2 g^2}{2} \delta
(\sum_i p^i) \int_0^\infty d \tau e^{\tau}
\Det \left (\frac{1 - \td X^2}{(1- \td V^2)^{13}} \right)\\
\times \exp \left(\half p^i_\mu Q^{ij}_{00} p^{j \mu} - p^i_\mu
Q^{ij}_{0 n} J^{j \mu}_n + \half J^{i}_{m \mu} Q^{ij}_{mn}J^{j
\mu}_{n} - \cJ_{cm}^i \cQ^{ij}_{mn} \cJ_{bn}^j \right).
\end{multline}
Here  $i, j \in {1, 2, 3, 4}$.  the matrices $\td V$ and $\td X$
are defined by
\begin{align}
\label{eq:vxtildedef} \td V_{mn} & = e^{-\frac{m}{2} \tau} V_{mn}
e^{-\frac{n}{2} \tau}, & \td X_{mn} & = e^{-\frac{m}{2} \tau}
X_{mn} e^{-\frac{n}{2} \tau}.
\end{align}
The matrices $Q^{ij}$  and $\cQ^{ij}$ are defined through the
tilded matrices $\td Q^{ij}$  and $\td \cQ^{ij}$
\begin{align}
\label{eq:qdef} \td Q^{ij}_{mn} &= e^{-\frac{m}{2}\tau}
Q^{ij}_{mn}e^{-\frac{n}{2}\tau},  & \td \cQ^{ij}_{mn} &=
e^{-\frac{m}{2}\tau} \cQ^{ij}_{mn}e^{-\frac{n}{2}\tau}
\end{align}
where the tilded matrices $\td Q$ and $\td \cQ$ are defined
through $\td V$, $\td U$, $\td X$
\begin{align}
\label{eq:qtildedef} \td Q^{\alpha\beta}   & = \td U^{\alpha 3}
\frac{1}{1 - \td V^2} \td V \td U^{3\beta}+ \td U^{\alpha\beta}, &
\td \cQ^{\alpha\beta}   & =  \td X^{\alpha 3} \frac{1}{1 - \td
X^2} \td X \td X^{3\beta}+ \td X^{\alpha\beta} , \nonumber \\
\td Q^{\alpha\alpha'}_{mn}   & = - \left(\td U^{\alpha 3}
\frac{1}{1 - \td V^2} C \td U^{3\alpha'}\right)_{mn} + \delta_{0m}
\delta_{0n} \tau, & \td \cQ^{\alpha\alpha'} & = - \td X^{\alpha 3}
\frac{1}{1 - \td X^2} C \td X^{3\alpha'}
\end{align}
with $\alpha,\beta  \in {1, 2} $;  $\alpha',\beta'  \in {3, 4} $.
The matrix $\td U$ includes zero modes while $\td V$ does not, so
one has to understand $\td U \td V$ in (\ref{eq:qtildedef}) as a
product of $\td U$, where the first column is dropped, and $\td
V$.  Similarly $\td V \td U$ is the product of $\td V$ and  $\td
U$ with the first row of $\td U$ omitted.

The matrices $Q^{ij}$ are not all independent for different $i$
and $j$.  The four-point amplitude is invariant under the twist
transformation  of either of the two vertices as well as under the
interchange of the two (see
Figure~\ref{f:quartic_diagrams}).  In addition the whole block matrix
$Q^{ij}_{mn}$ has been defined in such a way that it is symmetric
under the simultaneous exchange of $i$ with $j$ and $m$ with $n$.
Algebraically, we can use  properties (\ref{eq:nccycle},
\ref{eq:nctranspose}, \ref{eq:ncConjugate}) of Neumann
coefficients to show that the matrices $Q^{ij}$ satisfy
\begin{align}
(Q^{\alpha\beta})^T   & =  Q^{\beta\alpha},          & C
Q^{\alpha\beta} C   & =  Q^{3-\alpha \; 3-\beta},     &
Q^{\alpha\beta }      & =  Q^{\alpha+2 \;\beta+2},     \nonumber  \\
(Q^{\alpha'\beta'})^T & =  Q^{\beta'\alpha'},  & C
Q^{\alpha'\beta'}C  & =  Q^{7-\alpha' \;7-\beta'},   &
Q^{\alpha'\beta'}     & =  Q^{\alpha'-2 \;\beta'-2},   \\
(Q^{\alpha\alpha'})^T & =  Q^{\alpha'\alpha},    & C
Q^{\alpha\alpha'} C & =  Q^{3-\alpha \;7-\alpha'},   &
 Q^{\alpha\alpha'}    & =  Q^{\alpha+2 \;\alpha'-2}.  \nonumber
\end{align}
The analogous relations are satisfied by  ghost matrices $\cQ$.

Note that we still have some freedom in  the definition of the
zero modes of the matter matrices $Q$.  Due to the momentum
conserving delta function we can add to the exponent in the
integrand of (\ref{eq:4genfuncprelim}) any expression proportional
to $\sum p_i$.  To fix this freedom we require that after the
addition of such a term the new matrices $\bar Q$ satisfy $\bar
Q^{ii}_{00} = \bar Q^{ii}_{0n} = 0$.  This gives
\begin{align}
\label{eq:qbardef} \bar Q^{ij}_{00} & =  Q^{ij}_{00} - \half
Q^{jj}_{00} - \half Q^{ii}_{00} , & \bar Q^{ij}_{0n} & =
Q^{ij}_{0n} - Q^{jj}_{0n}.
\end{align}
and $\bar Q^{ij}_{mn} = Q^{ij}_{mn}$ for $m, n > 0$.  The addition
of any term proportional to $\sum p_i$ corresponds  in  coordinate
space to the addition of a total derivative.  In coordinate space
we have essentially integrated by parts the terms $\pd_\sigma
\pd^\sigma \psi_{\mu_1  \cdots  \mu_n}(x)$ and $\pd^{\mu_j}
\psi_{\mu_1  \cdots \mu_j \cdots \mu_n}(x)$ thus fixing the freedom
of integration by parts.

To summarize, we have rewritten $\Omega_4$ in terms of $\bar Q$'s
as
\begin{multline}
\label{eq:4genfunc} \Omega_4 = \frac{\cN^2 g^2}{2} \delta (\sum_i
p^i) \int_0^\infty d \tau e^{\tau}
\Det \left (\frac{1 - \td X^2}{(1- \td V^2)^{13}} \right)\\
\times \exp \left(\half p^i_\mu \bar Q^{ij}_{00} p^{j \mu} -
p^i_\mu \bar Q^{ij}_{0 n} J^{j \mu}_n + \half J^{i}_{m \mu} \bar
Q^{ij}_{mn}J^{j \mu}_{n}
 -\cJ_{cm}^i \cQ^{ij}_{mn} \cJ_{bn}^j \right).
\end{multline}
There are only three independent matrices $\bar Q$.  For  later
use we find it convenient to denote the independent $\bar Q$'s by
$A = \bar Q^{12}$,   $B = \bar Q^{13}$, $C = \bar Q^{14}$.  Then
the matrix $\bar Q^{ij}_{mn}$ can be written as
\begin{equation}
\bar Q^{i j}_{mn} =
\begin{pmatrix}
\label{eq:ABCdef}
       0            &     A_{mn}        &   B_{mn}             &     C_{mn}         \\
(-1)^{m+n}A_{mn}    &     0             & (-1)^{m+n} C_{mn}    &  (-1)^{m+n} B_{mn} \\
    B_{mn}          &     C_{mn}        &      0               &     A_{mn}         \\
 (-1)^{m+n} C_{mn}  & (-1)^{m+n} B_{mn} & (-1)^{m+n}A_{mn}     &       0
\end{pmatrix}.
\end{equation}
In the next section we derive off-shell amplitudes for the
massless fields by differentiating $\Omega_4$.  The generating
function $\Omega_4$ defined in (\ref{eq:4genfunc}) and
supplemented with the definition of the matrices $\td V$, $\td X$,
$\bar Q$, $\cQ$ given in (\ref{eq:udef}), (\ref{eq:vxtildedef}),
(\ref{eq:qdef}), (\ref{eq:qtildedef}), (\ref{eq:qbardef}) and
(\ref{eq:ABCdef}) provides us  with all  information about the
four-point tree-level off-shell amplitudes.

\subsection{Effective action for  massless fields}
\label{sec:masslessamps} 

In this subsection we compute explicit expressions for the general
quartic off-shell amplitudes of the massless fields, including
derivatives to all orders.
Our notation for the massless fields is, as
in (\ref{eq:stringfields}),
\begin{equation}
\ket{\Phi_{\text{massless}}} = \int d^d p \left(A_\mu(p) a_{-1}^\mu 
- i \alpha(p) b_{-1} c_{0}\right) \ket{p}.
\end{equation}
External states with $A_\mu$ and $\alpha$ in the $k$'th Fock space
are inserted using
\begin{align}
\label{eq:diffops} D^{A, k}_{\mu} = \int d p A_{\mu}(p) \frac{\pd
}{\pd J^{k}_{1\mu}}\Bigg |_{J^k = \cJ_{b, c}^k = 0} && \text{and}
&& D^{\alpha,k} = -i \int d p\, \alpha(p) \frac{\pd}{\pd \cJ^k_{b
1}} \frac{\pd}{\pd \cJ^k_{c 0}} \Bigg |_{J^k = \cJ_{b, c}^k  = 0}.
\end{align}
We can compute all quartic terms in the effective action
$\check{S}[A_\mu, \alpha]$ by computing quartic off-shell amplitudes
for the massless fields by acting on $\Omega_4$ with $D^A$ and
$D^\alpha$.  First consider the quartic term with four external $A$'s.
The relevant off-shell amplitude is given by $\prod_{i=1}^4 D^{A,
i}_{\mu_i} \Omega_4$ where $\Omega_4$ is given in (\ref{eq:4genfunc})
and $D^{A, i}_{\mu_i}$ is given in (\ref{eq:diffops}).  Performing the
differentiations we get
\begin{multline}
\label{eq:4GB} S_{A^4} =  \frac{1}{2}\, \cN^2 g^2 \int \prod_{i} d
p^i \delta\left(p^1+p^2+p^3+p^4 \right)
A^{\mu_1}(p_1) A^{\mu_2}(p_2) A^{\mu_3}(p_3) A^{\mu_4}(p_4) \\
\times \int_0^\infty d\tau e^{\tau}
 \Det \left (\frac{1 - \td X^2}{(1- \td V^2)^{13}} \right)
\Bigl (\cI^0_{A^4} + \cI^2_{A^4} +  \cI^4_{A^4} \Bigr) \exp
\left(\half p^i_\mu \bar Q^{ij}_{00} p^{j \mu} \right).
\end{multline}
Here $\mathcal{I}^0_{A^4}$, $\mathcal{I}^2_{A^4}$,
$\mathcal{I}^4_{A^4}$ are defined by
\begin{align}
\label{eq:4GBsuppl} \cI_{A^4}^0 = &\frac{1}{8} \sum_{i_i \neq i_j}
 \bar Q^{i_1 i_2}_{11} \bar Q^{i_3 i_4}_{11}
\eta_{\mu_{i_1} \mu_{i_2}} \eta_{\mu_{i_3} \mu_{i_4}} ,
\nonumber \\
\cI_{A^4}^2 = & \frac{1}{4} \sum_{i_i \neq  i_j}
 \bar Q^{i_1 i_2}_{11} \bar Q^{i_3 j_1}_{10}  \bar Q^{i_4 j_2}_{10}
 p^{j_1}_{\mu_{i_3}} p^{j_2}_{\mu_{i_4}}
 \eta_{\mu_{i_1} \mu_{i_2}}, \\
\cI_{A^4}^4 = &  \bar Q^{1i}_{10} \bar Q^{2j}_{10} \bar
Q^{3k}_{10} \bar Q^{4l}_{10}
           p^i_{\mu_1} p^j_{\mu_2} p^k_{\mu_3} p^l_{\mu_4}.
\nonumber
\end{align}
Other amplitudes with  $\alpha$'s and  $A$'s all have the same
pattern as (\ref{eq:4GB}).  The amplitude with one $\alpha$ and
three $A$'s is obtained by replacing $A_{\mu_{i_1}}(p^{i_1})$ in
formula (\ref{eq:4GB}) with $i\alpha(p^{i_1})$ and the sum of
$\cI_{A^4}^{0,2,4}$ with the sum of
\begin{align}
\label{eq:alpha3GB} \cI_{\alpha A^3}^1 & =  \half \sum_{i_i \neq
i_j} \cQ^{i_1 i_1}_{01} \bar Q^{i_2 i_3}_{11} \bar Q^{i_4k}_{10}
p^k_{\mu_{i_4}}   \eta_{\mu_{i_2} \mu_{i_3}},
  \nonumber \\
\cI_{\alpha A^3}^3 & =  \frac{1}{6} \sum_{i_i\neq i_j}
\cQ^{i_1i_1}_{01} \bar Q^{i_2 j}_{10} \bar Q^{i_3 k}_{10}  \bar
Q^{i_4 l}_{10}
 p^j_{\mu_{i_2}} p^k_{\mu_3} p^l_{\mu_4} .
\end{align}
The amplitude with two $A$'s and two $\alpha$'s is obtained by
replacing $A_{\mu_{i_1}}(p^{i_1}) A_{\mu_{i_2}}(p^{i_2})$ with $-
\alpha(p^{i_1}) \alpha(p^{i_2})$ and the sum of
$\cI_{A^4}^{0,2,4}$ with the sum of
\begin{align}
\label{eq:2alphas2GB} \cI_{\alpha^2 A^2}^0 & = \quarter \sum_{i_i
\neq i_j} (\cQ^{i_1i_1}_{01} \cQ^{i_2i_2}_{01} - \cQ^{i_1i_2}_{01}
\cQ^{i_2i_1}_{01})
   \bar Q^{i_3 i_4}_{11} \eta_{\mu_{i_3} \mu_{i_4}},
\nonumber \\
\cI_{\alpha^2 A^2}^2 &= \quarter \sum_{i_i \neq i_j}
(\cQ^{i_1i_1}_{01} \cQ^{i_2i_2}_{01} - \cQ^{i_1i_2}_{01}
\cQ^{i_2i_1}_{01}) \bar Q^{i_3 k}_{10}  \bar Q^{i_4 l}_{10}
p^k_{\mu_3} p^l_{\mu_4}.
\end{align}
It is straightforward to write down the analogous expressions for the
terms of order $\alpha^3A$ and $\alpha^4$.  However, as we shall see
later, it is possible to extract all the information about the
coefficients in the expansion of the effective action for $A_\mu$ in
powers of field strength up to $F^4$ from the terms of order $A^4, A^3
\alpha$, and $A^2 \alpha^2$.

The off-shell amplitudes (\ref{eq:4GB}), (\ref{eq:4GBsuppl}),
(\ref{eq:alpha3GB}) and (\ref{eq:2alphas2GB}) include contributions
from the intermediate gauge field.  To compute the quartic terms in the
effective action we must subtract, if nonzero, the amplitude with
intermediate $A_\mu$.  In the case of the abelian theory this
amplitude vanishes due to the twist symmetry.  In the nonabelian case,
however, the amplitude with intermediate $A_\mu$ is nonzero.  The
level truncation method in the next section makes it easy to
subtract this contribution at the stage of numerical computation.

\junk{As we see from  (\ref{eq:4GB}) the  full amplitude contains
infinite number of derivatives.  The gauge invariance is not
explicit since the amplitude is written in terms of potentials and
not the field strengths.  The situation is complicated by the fact
that one needs to perform the field redefinition to a new variable
$\hat A_\mu$ in order to write the action as a function of the
ordinary field strength.  For this reason even in the abelian case
the gauge invariance involves cancellations between different
amplitudes and it is not possible to write (\ref{eq:4GB}) in terms
of the ordinary field strength.  To exhibit the gauge invariance
and to compare the effective action computed here with the known
results obtained by other methods we expand the amplitudes in
powers of $p$, split summands that lead to distinct terms in the
action, find a field redefinition to the variable $\hat A_\mu$
that transforms according to the standard Yang-Mills
transformation law and write the result in the coordinate space in
terms of $\hat F_{\mu\nu}$.  Since the number of derivatives in all
the terms in the Born-Infeld side is less or equal to the number
of fields to identify terms that come from the Born-Infeld action
it will be sufficient to do the expansion keeping terms up to
$p^4$.

Luckily, the complications related to the field redefinition do not
involve terms that contribute to the Yang-Mills/Maxwell action.}  As
in (\ref{eq:expansion}), we expand the effective action in powers of
$p$.  As an example of a particular term appearing in this expansion,
let us consider the space-time independent (zero-derivative) term of
(\ref{eq:4GB}).  In the abelian case there is only one such term:
$A_\mu A^\mu A_\nu A^\nu$.  The coefficient of this term is
\begin{equation}
\label{eq:ceff0D} \gamma = \frac{1}{2}\, \cN^2 g^2 \int_0^\infty
d\tau e^{\tau} \Det \left (\frac{1 - \td X^2}{(1- \td V^2)^{13}}
\right) (A_{11}^2+B_{11}^2+C_{11}^2)
\end{equation}
where the matrices $A, B$ and $C$ are those in (\ref{eq:ABCdef}).
In the nonabelian case there are two terms, $\Tr\left(A_\mu A^\mu
A_\nu A^\nu\right)$ and $\Tr\left(A_\mu A_\nu A^\mu A^\nu\right)$,
which differ in the order of gauge fields.  The coefficients of
these terms are obtained by keeping $A_{11}^2 + C_{11}^2$ and
$B_{11}^2$ terms in $(\ref{eq:ceff0D})$ respectively.  

\subsection{Level truncation}
\label{sec:truncation}

Formula (\ref{eq:ceff0D}) and analogous formulae for the
coefficients of other terms in the effective action contain
integrals over complicated functions of infinite-dimensional
matrices.  Even after truncating the matrices to finite size,
these integrals are rather difficult to compute.  To get numerical
values for the terms in the effective action, we need a good
method for approximately evaluating integrals of the form
(\ref{eq:ceff0D}).  In this subsection we describe the method we
use to approximate these integrals.  For the four-point functions,
which are the main focus of the computations in this paper, the
method we use is equivalent to truncating the summation over
intermediate fields at finite field level.  Because the
computation is carried out in the oscillator formalism, however,
the complexity of the computation only grows polynomially in the
field level cutoff.

Tree diagrams with four external fields have a single internal
propagator with Schwinger parameter $\tau$.  It is convenient to do
a change of variables
\begin{equation}
\sigma = e^{-\tau}.
\end{equation}
We then truncate all matrices to size $L \times L$ and expand the
integrand in powers of $\sigma$ up to $\sigma^{M-2}$, dropping all
terms of higher order in $\sigma$.  We denote this approximation
scheme by $\{L, M\}$.  The $\sigma^n$ term of the series contains
the contribution from all intermediate fields at level $k = n+2$,
so in this approximation scheme we are keeping all oscillators
$a^{\mu}_{k\le L}$ in the string field expansion, and all
intermediate particles in the diagram of mass $m^2 \le M - 1$.  We
will use the approximation scheme $\{L, L\}$ throughout this
paper.  This approximation really imposes only one restriction---
the limit on the mass of  the intermediate particle.  It is perhaps
useful to compare the approximation scheme we are using here with
those used in previous work on related problems.  In
\cite{Taylor-amplitudes} analogous integrals were computed by
numerical integration.  This corresponds to $\{L, \infty\}$
truncation.  In earlier papers on level truncation in string field
theory, such as \cite{Kostelecky-Samuel, Sen-Zwiebach,
Moeller-Taylor} and many others, the $(L, M)$ truncation scheme
was used, in which fields of mass up to $L - 1$ and interaction
vertices with total mass of fields in the vertex up to $M - 3$ are
kept.  Our $\{L, L\}$ truncation scheme is equivalent to the $(L,
L+2)$ truncation scheme by that definition.

To explicitly  see how the $\sigma$ expansion works let us write
the expansion in $\sigma$ of a generic  integrand and take the
integral term by term
\begin{equation}
\label{eq:poles} \int_0^1 \frac{d\sigma}{\sigma^2} \sigma^{p^2}
\sum_{n=0}^\infty c_n(p^i) \sigma^n = \sum_{n=0}^\infty
\frac{c_n(p^i)}{p^2+n-1}.
\end{equation}
Here $p = p_1 + p_2 = p_3 +p_4$ is the intermediate momentum.  This
is the expansion of the amplitude into poles corresponding to the
contributions of (open string) intermediate particles of fixed
level.  We can clearly see that dropping higher powers of $\sigma$
in the expansion means dropping the contribution of very massive
particles.  We also see that to subtract the contribution from the
intermediate fields $A_\mu$ and $\alpha$ we can simply omit the
term $c_1(p) \sigma^{p^2-1}$ in (\ref{eq:poles}).

While the Taylor expansion of the integrand might seem difficult,
it is in fact quite straightforward.  We notice that $\td V^{rs}$,
and $\td X^{rs}$ are both of order $\sigma$.  Therefore we can
simply expand the integrand in powers of matrices $\td V$ and $\td
X$.  For example, the determinant of the matter Neumann
coefficients is
\begin{equation}
\Det (1-\td V^2)^{- 13}= \exp\left(-13 \Tr\Log(1 - \td V^2)
\right).
\end{equation}
Looking again at (\ref{eq:4GB}) we notice that the only matrix
series' that we will need are $\Log(1 - \td V^2)$ for the
determinant (and the analogue for $\tilde{X}$) and $1/(1 - \td
V^2)$ for $\bar Q^{ij}$.  Computation of these series is
straightforward.

It is also easy to estimate how computation time grows with $L$
and $M$.  The most time consuming part of the Taylor expansion in
$\sigma$ is the matrix multiplication.  Recall that $\td V$ is an
$L\times L$ matrix whose coefficients are proportional to
$\sigma^n$ at leading order.  Elements of $\td V^k$ are polynomials
in $\sigma$ with $M$ terms.  To construct a series $a_0 + a_1 \td
V + \cdots + a_M \td V^M+ \cO(\sigma^{M+1})$ we need $M$ matrix
multiplications $\td V^k \cdot \td V$.  Each matrix multiplication
consists of $L^3$ multiplications of its elements.  Each
multiplication of the elements has on the average $M/2$
multiplications of monomials.  The total complexity therefore
grows as $L^3 M^2$.

The method just described allows us to compute approximate
coefficients in the effective action at any particular finite level of
truncation.  In \cite{Taylor-amplitudes}, it was found empirically
that the level truncation calculation gives approximate results for
finite on-shell and off-shell amplitudes with errors which go as a
power series in $1/L$.  Based on this observation, we can perform a
least-squares fit on a finite set of level truncation data for a
particular term in the effective action to attain a highly accurate
estimate of the coefficient of that term.  We use this method to
compute coefficients of terms in the effective action which are
quartic in $A$ throughout the remainder of this paper.

\section{The Yang-Mills action}
\label{sec:Yang-Mills} In this section we assemble the Yang-Mills
action, picking the appropriate terms from the two, three and
four-point Green functions.  We write the Yang-Mills action as
\begin{multline}
S_{YM} =  \int d^d x {\rm Tr} \Bigl(-\frac{1}{2} \pd_\mu A_\nu \pd
^\mu A^\nu + \frac{1}{2} \pd_\mu A_\nu \pd^\nu A^\mu \\ + i g_{YM}
\pd_\mu A_\nu[A^\mu, A^\nu] + \quarter g_{YM}^2 [A_\mu, A_\nu]
[A^\mu, A^\nu] \Bigr ).  \label{eq:Yang-Mills}
\end{multline}
In section \ref{sec:YMquadratic} we consider the quadratic terms
of the Yang-Mills action.  In section \ref{sec:YMcubic} 
consider the cubic terms and identify the Yang-Mills
coupling constant $g_{YM}$ in terms of the SFT (three tachyon)
coupling constant $g$.  This provides us with the expected value
for the quartic term.  In section \ref{sec:YMquartic} we present
the results of a numerical calculation of the (space-time
independent) quartic terms and verify that we indeed get the
Yang-Mills action.

\subsection{Quadratic terms}
\label{sec:YMquadratic} The quadratic term in the action for
massless fields, calculated from (\ref{eq:action}), and
(\ref{eq:v2}) is
\begin{equation} \check S_{A^2} = \int d^d x \Tr \Bigl(-\frac{1}{2} \pd_\mu
A_\nu\, \pd^\mu A^\nu - \alpha^2 + \sqrt{2} \alpha\, \pd_\mu A^\mu
\Bigr).  \label{eq:quadraticmassless}
\end{equation}
Completing the square in $\alpha$ and integrating the term $(\pd
A)^2$ by parts we obtain
\begin{equation}
\check S_{A_2} = \int d^d x {\rm Tr} \Bigl (-\frac{1}{2} \pd_\mu
A_\nu \,\pd ^\mu A^\nu + \frac{1}{2} \pd_\mu A_\nu\, \pd^\nu A^\mu
- B^2 \Bigr ) \label{eq:YMquadratic}
\end{equation}
where we denote 
\begin{equation}
B = \alpha - \frac{1} {\sqrt{2}} \pd_ \mu
A^\mu\,.
\end{equation}
Eliminating $\alpha$ using the leading-order equation of motion, $B =
0$, leads to the quadratic terms in (\ref{eq:Yang-Mills}).
Subleading terms in the equation of motion for $\alpha$ lead to
higher-order terms in the effective action, to which we return in the
following sections.
\junk{and the terms of order $\cO(A^2)$ and $\cO(\alpha^2)$ are absent.
If we eliminate $\alpha$ from (\ref{eq:YMquadratic}) using this
equation of motion we will have the quadratic part of the
Yang-Mills action plus terms of order $\pd^4 A^4$.  In the abelian case the
cubic terms (\ref{eq:cubicAa2}) vanish.  Eliminating $\alpha$ from
(\ref{eq:YMquadratic}) in the abelian theory one gets the Maxwell
action plus terms of order $A^6$.}

\subsection{Cubic terms}
\label{sec:YMcubic} The cubic terms in the action for the massless
fields are obtained by differentiating (\ref{eq:Omega3}).  The terms
cubic in $A$ are given by
\begin{multline}
\check S_{A^3} = \frac{\cN g}{3} \int \prod_i d p_i
\delta(\sum_j p_j) \Tr\Bigl(A_\mu(p_1) A_\nu(p_2)
A_\lambda(p_3)\Bigr)
\exp\bigl(\half p^{r} V^{rr}_{00} p^{r}\bigr) \\
\times \Bigl( \bigl ( \eta^{\nu\lambda} p^{r \mu} V^{r1}_{01}
V^{32}_{11} + \eta^{\mu\lambda} p^{r \nu} V^{r2}_{01} V^{13}_{11}
+ \eta^{\mu\nu} p^{r \lambda} V^{r3}_{01} V^{12}_{11} \bigr ) \\
+  p^{r \mu} V^{r1}_{01} p^{s \nu} V^{s2}_{01} p^{t \lambda}
V^{t3}_{01} \Bigr).
\end{multline}
To compare with the Yang-Mills action we perform a Fourier
transform and use the properties of the Neumann coefficients to
combine similar terms.  We then get
\begin{multline}
\check S_{A^3} = - i \cN g \int d x \Tr \Bigl (V^{12}_{11}
V^{12}_{01}\bigl( \pd_\mu \td A_\nu [\td A^\mu, \td A^\nu] \bigr )
\\ + \frac{1}{3} (V^{12}_{01})^3 \bigl(\pd_\lambda \td A^\mu
\pd_\mu \td A^\nu \pd_\nu \td A^\lambda - \pd_\nu \td A^\mu
\pd_\lambda \td A^\nu \pd_\mu \td A^\lambda \bigr) +
\bigl(V_{01}^{12}\bigr)^3 [\td A_\nu, \pd^\ld \td A_\mu] \pd^{\mu}
\pd^{\nu} \td A_\ld \Bigr) \label{eq:cubicmassless}
\end{multline}
where, following the notation introduced in (\ref{eq:tilde}), we have
\begin{equation}
\td A_\mu = \exp(-\half V^{11}_{00} \pd^2) A_\mu.
\end{equation}
To reproduce the cubic terms in the Yang-Mills action, we are
interested in the terms in (\ref{eq:cubicmassless}) of order $\partial
A^3$.  The remaining terms and the terms coming from the expansion of
the exponential of derivatives contribute to higher-order terms in the
effective action, which we discuss later.  The cubic terms in the
action involving the $\alpha$ field are
\begin{align}
\check S_{A\alpha^2} & = - i \cN g V_{01}^{12} \left(
X_{01}^{12}\right)^2 \int d x  \Tr\bigl(\td A^\mu \left[ \pd_\mu
\td \alpha , \td
\alpha \right] \bigr) \label{eq:cubicAa2}, \\
\check S_{A^2\alpha} & = \check S_{\alpha^3} = 0 .  \nonumber
\end{align}
$\check S_{A^2\alpha}$ vanishes  because $X^{11}_{01} = 0$,
and $\check S_{\alpha^3}$ is zero because $[\alpha, \alpha] = 0$.
After $\alpha$ is eliminated using its equation of motion,
(\ref{eq:cubicAa2}) first contributes terms at order $\pd^3 A^3$.

The first line of (\ref{eq:cubicmassless}) contributes to the cubic
piece of the $F^2$ term.  Substituting the explicit values of the
Neumann coefficients:
\begin{eqnarray}
V^{11}_{00} = -\log(27/16), & \hspace*{0.8in} & V^{12}_{11} =
16/27,\\
V^{12}_{01} = - 2 \sqrt{2}/3\sqrt{3}, &  & X^{12}_{01}= 4/(3\sqrt{3}).
\nonumber
\end{eqnarray}
we write the lowest-derivative term of (\ref{eq:cubicmassless}) as
\begin{equation}
S^{[1]}_{A^3} = i \frac{g}{\sqrt{2}} \int d^d x {\rm
Tr}\left(\pd_\mu A_\nu[ A^\mu, A^\nu]\right)\,. \label{eq:YMcubic}
\end{equation}

We can now predict the value of the quartic amplitude at zero
momentum.  {}From (\ref{eq:Yang-Mills}) and
(\ref{eq:YMcubic}) we see that the Yang-Mills constant is related
to the SFT coupling constant by
\begin{equation}
g_{YM} = \frac{1}{\sqrt{2}} g.
\end{equation}
This is the same relation between the gauge boson and tachyon
couplings as the one given in formula (6.5.14) of Polchinski
\cite{Polchinski}.  We expect the nonderivative part of the quartic
term in the effective action to add to the quadratic and cubic terms
to form the full Yang-Mills action, so that
\begin{equation}
S^{[0]}_{A_4} = \quarter g_{YM}^2 [A_\mu, A_\nu]^2.
\label{eq:S4exp}
\end{equation}

\subsection{Quartic terms}
\label{sec:YMquartic} As we have just seen, to get the full
Yang-Mills action the quartic terms in the effective action at
$p=0$ must take the form (\ref{eq:S4exp}).  We write the
nonderivative part of the SFT quartic effective action as
\begin{equation}
S_{A^4}^{[0]} =  g^2 \int d x \Big(\gamma_{+}{\rm Tr}(A_\mu
A^\mu)^2 + \quarter \gamma_{-}{\rm Tr}[A_\mu, A^\nu]^2\Big).
\label{eq:quarticterm}
\end{equation}
We can use the method described in section \ref{sec:truncation} to
numerically approximate the coefficients $\gamma_+$ and $\gamma_-$ in
level truncation.
In the limit $L \rightarrow \infty$ we expect  that $\gamma_+
\rightarrow 0$ and that $\gamma_- \rightarrow g_{YM}^2/ g^2 =
1/2$.  As follows from formula (\ref{eq:ceff0D}) and the comment
below it $\gamma_{\pm}$ are given by:
\begin{align}
\gamma_{+} & = \half \cN^2 \int_0^\infty e^\tau d\tau  \Det \left
(\frac{1 - \td X^2}{(1- \td V^2)^{13}} \right)
(A_{11}^2 + B_{11}^2 + C_{11}^2), \nonumber \\
\gamma_{-} & = \cN^2 \int_0^\infty e^\tau d\tau  \Det \left
(\frac{1 - \td X^2}{(1- \td V^2)^{13}} \right) B_{11}^2.
\end{align}
We have calculated these integrals including contributions from the
first $100$ levels.  We have found that as the level $L$ increases the
coefficients $\gamma_+$ and $\gamma_-$ indeed converge to their
expected values \footnote{We were recently informed of an analytic
proof of this result in SFT, which will appear in
\cite{Berkovits-Schnabl}}.  The leading term in the deviation decays
as $1/L$ as expected.  Figure \ref{f:quartic} shows the graphs of
$\gamma_\pm(L)$ vs $L$.  \FIGURE{\includegraphics{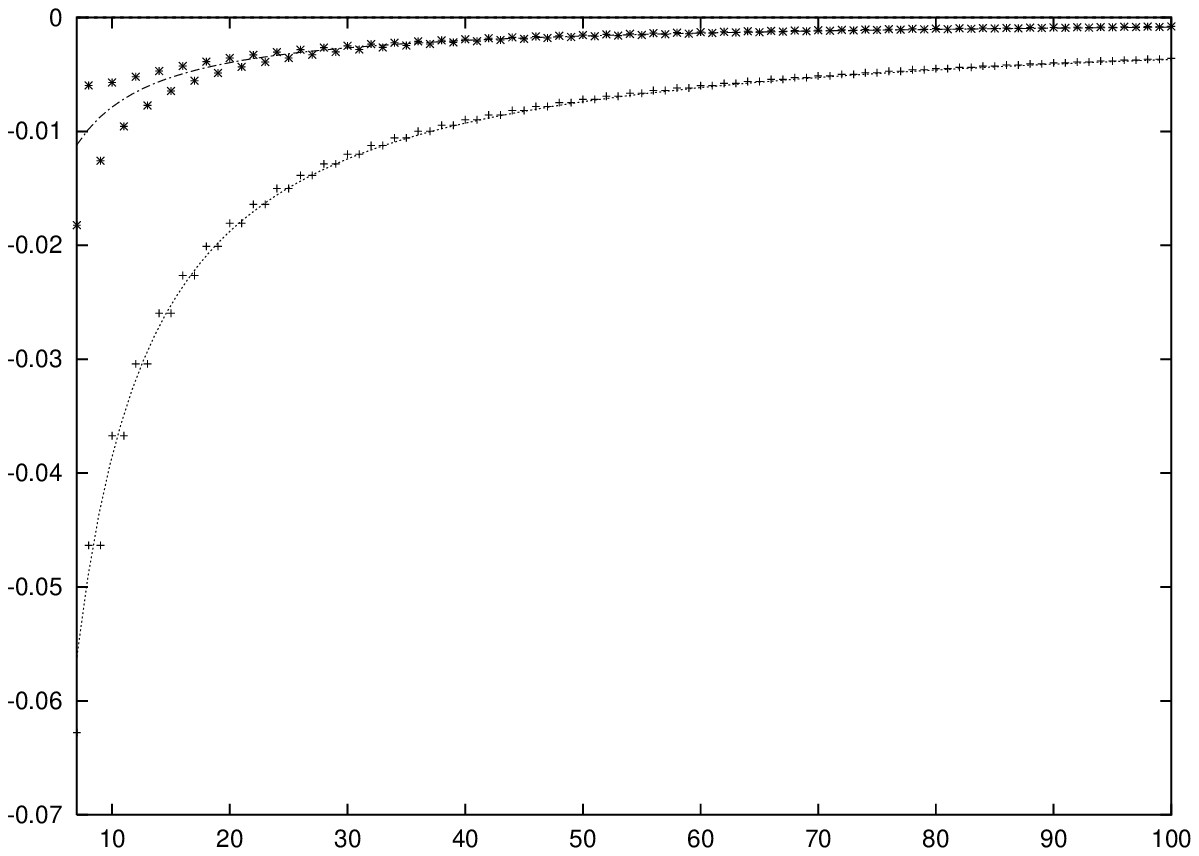}
\caption[x]{\footnotesize Deviation of the coefficients of quartic
terms in the effective action from the expected values, as a function
of the level of truncation $L$.  The coefficient $\gamma_+$ is shown
with crosses and $\gamma_- -1/2$ is shown with stars.  The curves
given by fitting with a power series in $1/L$ are graphed in both cases.}
\label{f:quartic}}

\noindent Table \ref{t:quartic} explicitly lists the results from
the first 10 levels.  
\begin{table}[htp]
\centering
\begin{tabular} {|l|c|c|r|}
\hline
Level & $\gamma_{+}(n)$ &  $\gamma_{-}(n)$ & $\gamma_{-}(n) - \half$ \\
\hline
0  &  -0.844 \ &  0      & -0.500 \ \ \\
\hline
2  &  -0.200  &  0.592 \ &  0.092  \ \ \\
\hline
3  &  -0.200  &  0.417  &  -0.083 \ \ \\
\hline
4  &  -0.097  &  0.504  &  0.004  \ \ \\
\hline
5  &  -0.097  &  0.468  & -0.032  \ \ \\
\hline
6  &  -0.063  &  0.495  & -0.005  \ \ \\
\hline
7  &  -0.063  &  0.483  & -0.017  \ \ \\
\hline
8  &  -0.047  &  0.494  & -0.006  \ \ \\
\hline
9  &  -0.047  &  0.487  & -0.013  \ \ \\
\hline
10 &  -0.037  &  0.494  & -0.006  \ \ \\
\hline
\end{tabular}
\caption{\footnotesize Coefficients of  the constant quartic terms
in the action for the first 10 levels.} \label{t:quartic}
\end{table}
At level $100$ we get $\gamma_{+} = 0.0037$,  $\gamma_{-} = 0.4992$
which is within $0.5\%$  of the expected values.  One can improve
precision even more by doing a least-squares fit of
$\gamma_{\pm}(L)$ with an expansion in powers of $1/L$ with
indeterminate coefficients.  The contributions to $\gamma_\pm$ from
the even and odd level fields are oscillatory.  Thus, the fit for
only even or only odd levels works much better.  The least-squares
fit for the last 25 even levels gives
\begin{align}
\gamma_+(L) & \approx  - 5 \cdot{10}^{-8}- \frac{0.35807}{L} -
\frac{0.0091}{L^2}- \frac{1.6}{L^3}  + \frac{15}{L^4} + \cdots \nonumber \\
\gamma_-(L) & \approx \half - 2 \cdot{10}^{-8}- \frac{0.0795838}{L}
+ \frac{0.1212}{L^2} + \frac{1.02}{L^3} - \frac{1.24}{L^4} + \cdots.
\end{align}
We see that when $L \rightarrow \infty$ the fitted values of
$\gamma_{\pm}$ are in agreement with the Yang-Mills quartic term
to 7 digits of precision \footnote{Note that in \cite{Taylor-action}, an
earlier attempt was made to calculate the coefficients $\gamma_{\pm}$
from SFT.  The results in that paper are incorrect; the error made
there was that odd-level fields, which do not contribute in the
abelian action due to twist symmetry, were neglected.  As these fields
do contribute in the nonabelian theory, the result for $\gamma_-$
obtained in \cite{Taylor-action} had the wrong numerical value.  Our
calculation here automatically includes odd-level fields, and
reproduces correctly the expected value.}.  

The calculations we have described so far provide convincing evidence
that the SFT effective action for $A_\mu$ reproduces the nonabelian
Yang-Mills action.  This is encouraging in several respects.  First, it
shows that our method of computing Feynman diagrams in SFT is working
well.  Second, the agreement with on-shell calculations is another
direct confirmation that cubic SFT provides a correct off-shell
generalization of bosonic string theory.  Third, it encourages us to
extend these calculations further to get more information about the
full effective action of $A_\mu$.

\section{The abelian Born-Infeld action}

In this section we consider the abelian theory, and compute terms
in the effective action which go beyond the leading Yang-Mills
action computed in the previous section.  As discussed in Section
2.3, we expect that the effective vector field theory computed
from string field theory should be equivalent under a field
redefinition to a theory whose leading terms at each order in $A$
take the Born-Infeld form (\ref{eq:BIexpansion}).  In this section
we give evidence that this is indeed the case.  In the abelian
theory, the terms at order $A^3$ vanish identically, so the
quartic terms are the first ones of interest beyond the quadratic
Yang-Mills action.  In subsection \ref{sec:abelian2der} we use our
results on the general quartic term from \ref{sec:masslessamps} to
explicitly compute the terms in the effective action at order
$\partial^2 A^4$.  We find that these terms are nonvanishing.  We
find, however, that the gauge invariance of the effective action
constrains the terms at this order to live on a one-parameter
family of terms related through field redefinitions, and that the
terms we find are generated from the Yang-Mills terms $\hat{F}^2$
with an appropriate field redefinition.  We discuss general issues
of field redefinition and gauge invariance in subsection
\ref{sec:fieldredefinitions}; this discussion gives us a framework
with which to analyze more complicated terms in the effective
action.  In subsection \ref{sec:abelian4der} we analyze terms of
the form $\partial^4A^4$, and show that these terms indeed take
the form predicted by the Born-Infeld action after the appropriate
field redefinition.  In subsection \ref{sec:a-higher} we consider
higher-order terms with no derivatives, and give evidence that
terms of order $(A \cdot A)^n$ vanish up to $n = 5$ in the string
field theory effective action.

\subsection{Terms of the form $\partial^2 A^4$}
\label{sec:abelian2der} In the abelian theory, all terms in the
Born-Infeld action have the same number of fields and derivatives.
If we assume that the effective action for $A_\mu$ calculated in
SFT directly matches the Born-Infeld action (plus higher-order
derivative corrections) we would expect the $\pd^{2}A^{4}$ terms
in the expansion of the effective action to vanish.  The most
general form of the quartic terms with two derivatives is
parameterized as \footnote{Recall that in section
\ref{sec:4pointgenfunc} we fixed the integration by parts freedom
by integrating by parts all terms with $\pd^2 A_\ld$ and $\pd
\cdot A$.  Formula (\ref{eq:2der}) gives the most general
combination of terms with four $A$'s and two derivatives that do
not have $\pd^2 A_\ld$ and $\pd \cdot A$.}
\begin{multline}
\label{eq:2der} S_{A^4}^2 = g^2 \int d^{26} x  \bigl (  c_1 A_\mu
A^\mu  \pd_\sigma A_\nu \pd^\sigma A^\nu +  c_2 A_\mu A_\nu
\pd_\sigma A^\mu \pd^\sigma A^\nu +
 c_3 A_\mu A_\nu \pd^\mu A_\sigma \pd^\nu A^\sigma \\
 + c_4 A_\mu A_\nu \pd^\nu A_\sigma \pd^\sigma A^\mu +
 c_5 A_\mu A_\nu  A_\sigma \pd^\mu \pd^\nu A^\sigma +
 c_6 A_\mu A^\mu \pd_\sigma A^\nu \pd_\nu A^\sigma \bigr).
\end{multline}

When $\alpha$ is eliminated from the massless effective action
$\check{S}$ using the equation of motion, we might then expect that
all coefficients $c_n$ in the resulting action (\ref{eq:2der}) should
vanish.  Let us now compute these terms explicitly.  {}From
(\ref{eq:YMquadratic}) and (\ref{eq:cubicAa2}) we see that the
equation of motion for $\alpha$ in the effective theory of the massless
fields reads (in the abelian theory)
\begin{equation}
\label{eq:alphaEOM} \alpha =  \frac{1}{\sqrt{2}} \pd^\mu A_\mu +
\cO( (A, \alpha)^3) \,.
\end{equation}
The coefficients $c_1, \ldots, c_6$ thus get
contributions from the two-derivative term of (\ref{eq:4GB}), the
one-derivative term of (\ref{eq:alpha3GB}) and the zero-derivative
term of (\ref{eq:2alphas2GB}).  We first consider the contribution
from the four-gauge boson amplitude (\ref{eq:4GB}).  All the
expressions for these contributions, which we denote $(\delta c_i
)_{A^4}$, are of the form
\begin{equation}
(\delta c_i )_{A^4} = \frac{1}{2}\, \cN^2 \int_0^\infty d\tau
e^{\tau}
  \Det \left
(\frac{1 - \td X^2}{(1- \td V^2)^{13}} \right) P_{\pd^2 A^4, i}(A,
B,
  C).
\end{equation}
Here $P_{\pd^2 A^4, i}$ are polynomials in the elements of the
matrices $A$, $B$ and $C$ which were defined in (\ref{eq:ABCdef}).
It is straightforward to derive expressions for the polynomials
$P_{\pd^2 A^4, i}$ from (\ref{eq:4GB}) and (\ref{eq:4GBsuppl}), so
we just give the result here
\begin{align}
P_{\pd^2 A^4, 1} & = -2 \bigl (A_{11}^2 A_{00} + B_{11}^2 B_{00} +
C_{11}^2 C_{00}
\bigr),  \nonumber\\
P_{\pd^2 A^4, 2}& = -2 \bigl ( A_{11}^2 (B_{00} + C_{00}) +
B_{11}^2 (A_{00} + C_{00})
                                             +  C_{11}^2 (A_{00} + B_{00})
\bigr),  \nonumber\\
P_{\pd^2 A^4, 3} & =  2 \bigl ( A_{11}(B_{10}^2 + C_{10}^2) -
B_{11}(A_{10}^2 + C_{10}^2)
                                         +  C_{11}(A_{10}^2 + B_{10}^2)
\bigr),           \\
P_{\pd^2 A^4, 4} & =  4 \bigl ( A_{11} A_{10 }(B_{10} + C_{10}) -
B_{11} B_{10} (A_{10} + C_{10})
                                         +  C_{11} C_{10} (A_{10} + B_{10})
\bigr),  \nonumber\\
P_{\pd^2 A^4, 5} & =  4 \bigl ( A_{11} B_{10 } C_{10} - B_{11}
A_{10} C_{10} +
                                                       C_{11} A_{10} B_{10}
\bigr),  \nonumber\\
P_{\pd^2 A^4, 6} & =  2 \bigl ( A_{11} A_{10 }^2- B_{11} B_{10}^2
+ C_{11} C_{10}^2 \bigr).  \nonumber
\end{align}
The terms in the effective action $\check{S}$ which contain
$\alpha$'s and contribute to $S[A]$ at order $\partial^2 A^4$ can
similarly be computed from (\ref{eq:alpha3GB}) and are given by
(\ref{eq:2alphas2GB})
\begin{equation}
\label{eq:2derivatives} \check{S}_{\alpha A^3}^{[1]} + \check{
S}_{\alpha^2 A^2}^{[0]}
 = g^2 \int d^{26} x
\bigl ( \sigma_1 \alpha A_\mu A_\nu \pd^\mu A^\nu +  \sigma_2
\pd^\mu \alpha A_\mu A_\nu A^\nu + \sigma_3 \alpha^2 A_\mu A^\mu
\bigr )
\end{equation}
where the coefficients $\sigma_i$ are given by
\begin{align}
P_{\pd \alpha A^3, 1} &= 4\cQ^{11}_{01} \bigl (A_{11}
(B_{10}+C_{10}) -
                   B_{11} (A_{10}+C_{10}) + C_{11} (B_{10}+A_{10}) \bigr) , \nonumber \\
P_{\pd \alpha A^3, 2} &=  4\cQ^{11}_{01}
                \bigl (A_{11} A_{10} - B_{11} B_{10} + C_{11} C_{10} \bigr)  \\
P_{\alpha^2 A^2} \ \ &=  2 \bigl( (\cQ^{11}_{01})^2 -
(\cQ^{12}_{01})^2\bigr) A_{11} -
                    \bigl( (\cQ^{11}_{01})^2 -  (\cQ^{13}_{01})^2\bigr) B_{11} +
                    \bigl( (\cQ^{11}_{01})^2 -  (\cQ^{14}_{01})^2\bigr) C_{11}.  \nonumber
\end{align}
Computation of the integrals up to level $100$ and using a
least-squares fit gives us
\begin{align}
(\delta c_1)_{A^4} & \approx - 2.1513026 , & (\delta c_4)_{A^4} &
\approx \ 0.9132288  , &
\sigma_1 & \approx - 0.4673613  ,  \nonumber \\
(\delta c_2)_{A^4} & \approx  - 4.3026050 , & (\delta c_5)_{A^4} &
\approx  - 2.0134501, &
 \sigma_2 & \approx\ 0.2171165 ,           \\
(\delta c_3)_{A^4} & \approx - 2.0134501 ,  & (\delta c_6)_{A^4} &
\approx  \ 1.4633393 , & \sigma_3 & \approx \ 1.6829758
.\nonumber
\end{align}
Elimination of $\alpha$ with (\ref{eq:alphaEOM}) gives
\begin{align}
c_1 & \approx - 2.1513026,  &  c_4 & \approx 4.302605 , \nonumber \\
c_2 & \approx - 4.302605 ,  &  c_5 & \approx 0,            \label{eq:cs}\\
c_3 & \approx 0,  &  c_6 & \approx 2.1513026.  \nonumber
\end{align}
These coefficients are not zero, so that the SFT effective action
does not reproduce the abelian Born-Infeld action in a
straightforward manner.  Thus, we need to consider a field
redefinition to put the effective action into the usual
Born-Infeld form.  To understand how this field redefinition
works, it is useful to study the gauge transformation in the
effective theory.  Without directly computing this gauge
transformation, we can write the general form that the
transformation must take; the leading terms can be parameterized
as
\begin{multline}
\delta A_\mu =
\pd_{\mu}\ld+g_{YM}^{2}(\varsigma_{1}A^{2}\pd_{\mu}\ld+
\varsigma_{2}A_{\nu}\pd_{\mu}A^{\nu}\ld\\
+\varsigma_{3}A_{\nu}\pd^{\nu}A_{\mu}\ld+\varsigma_{4}A_{\mu}\pd\cdot
A\ld +\varsigma_{5}A_{\mu}A_{\nu}\pd^{\nu}\ld )+ {\cal
O}(\pd^{3}A^{2}\ld).  \label{eq:gauge}
\end{multline}
The action (\ref{eq:2der}) must be invariant under this gauge
transformation.  This gauge invariance imposes a number of {\it a
priori} restrictions on the coefficients $c_i$, $\varsigma_i$.  When
we vary the $F^2$ term in the effective action (\ref{eq:Yang-Mills}) the
nonlinear part of (\ref{eq:gauge}) generates $\pd^{3} A^{3}\ld$
terms.  Gauge invariance requires that these terms cancel the
terms arising from the linear gauge transformation of the $\pd^2
A^4$ terms in (\ref{eq:2der}).  This cancellation gives homogeneous
linear equations for the parameters $c_i$ and $\varsigma_i$.  The
general solution of these equations depends on one free parameter
$\gamma$:
\begin{align}
\label{eq:2dcoefficients} c_{1} &= - c_6 =  - \gamma  ,  &
\varsigma_{1} &=  - \gamma, \nonumber \\
c_2 & = - c_4 = - 2 \gamma, & \varsigma_5 &= - 2 \gamma, \\
 c_{3} &= c_5 = 0,  &   \varsigma_{2} &= \varsigma_{3} = \varsigma_{4} = 0.  \nonumber
\end{align}
The coefficients $c_i$ calculated above satisfy these relations to
7 digits of precision.  {}From the numerical values of the $c_i$'s,
we find
\begin{align}
\label{eq:gamma} \gamma \approx  2.1513026 \pm 0.0000005.
\end{align}

We have thus found that the $\partial^2 A^4$ terms in the effective
vector field action derived from SFT lie on a one-parameter family
of possible combinations of terms which have a gauge invariance of
the desired form.  We can identify the degree of freedom
associated with this parameter as arising from the existence of a
family of field transformations with nontrivial terms at order
$A^3$
\begin{align}
\label{eq:fieldred2}
& \hat A_{\mu}= A_{\mu} +  g^2 \gamma A^{2} A_{\mu}, \\
& \hat \ld = \ld. \nonumber
\end{align}
We can use this field redefinition to relate a field $\hat{A}$
with the standard gauge transformation $\delta \hat{A}_\mu =
\partial_\mu \hat{\lambda}$ to a field $A$ transforming under
(\ref{eq:gauge}) with $\varsigma_i$ and $\gamma$ satisfying
(\ref{eq:2dcoefficients}).  Indeed, plugging this change of
variables into
\begin{align}
\delta \hat A_\mu & = \pd \hat \ld,  \\
S_{BI} & = -\quarter \int d x \hat F^2 + {\cal O} (\hat{F}^3).  \nonumber
\end{align}
gives (\ref{eq:gauge}) and  (\ref{eq:2der}) with $c_i$,
$\varsigma_i$ satisfying (\ref{eq:2dcoefficients}).

We have thus found that nonvanishing $\partial^2A^4$ terms arise in
the vector field effective action derived from SFT, but that these
terms can be removed by a field redefinition.  We would like to
emphasize that the logic of this subsection relies upon using the fact
that the effective vector field theory has a gauge invariance.  The
existence of this invariance constrains the action sufficiently that
we can identify a field redefinition that puts the gauge
transformation into standard form, without knowing in advance the
explicit form of the gauge invariance in the effective theory.  Knowing
the field redefinition, however, in turn allows us to identify this
gauge invariance explicitly.  This interplay between field
redefinitions and gauge invariance plays a key role in understanding
higher-order terms in the effective action, which we explore further
in the following subsection.

\subsection{Gauge invariance and field redefinitions}
\label{sec:fieldredefinitions}
In this subsection we discuss some aspects of the ideas of gauge
invariance and field redefinitions in more detail.  In the previous
subsection, we determined a piece of the field redefinition relating
the vector field $A$ in the effective action derived from string field
theory to the gauge field $\hat{A}$ in the Born-Infeld action by using
the existence of a gauge invariance in the effective theory.  The
rationale for the existence of the field transformation from $A$ to
$\hat{A}$ can be understood based on the general theorem of the
rigidity of the Yang-Mills gauge transformation \cite{Henneaux-db,
Henneaux-review}.  This theorem states that any deformation of the
Yang-Mills gauge invariance can be mapped to the standard gauge
invariance through a field redefinition.  At the classical level this
field redefinition can be expressed as
\begin{align}
\hat{A}_\mu &= \hat{A}_\mu(A), \nonumber \\
\hat \ld &= \hat{\ld}(A, \ld).
\end{align}
This theorem explains, for example, why noncommutative Yang-Mills
theory, which has a complicated gauge invariance involving the
noncommutative star product, can be mapped through the Seiberg-Witten
map (field redefinition) to a gauge theory written in terms of a gauge
field with standard transformation rules \cite{Seiberg-Witten,
Grigoriev-Henneaux}.  Since in string field theory the parameter
$\alpha'$ (which we have set to unity)
parameterizes the deformation of the standard gauge
transformation of $A_\mu$, the theorem states that some field
redefinition exists which takes the effective vector field theory
arising from SFT to a theory which can be written in terms of the
field strength $\hat{F}_{\mu \nu}$ and covariant derivative $\hat{D}_\mu$ of
a gauge field $\hat{A}_\mu$ with the standard transformation
rule\footnote{In odd dimensions there would also be a possibility of
Chern-Simons terms}.

There are two ways in which we can make use of this theorem.  Given
the explicit expression for the effective action from SFT, one can
assume that such a transformation exists, write the most general
covariant action at the order of interest, and find a field
redefinition which takes this to the effective action computed in
SFT.  Applying this approach, for example, to the $\partial^2 A^4$
terms discussed in the previous subsection, we would start with
the covariant action $\hat{F}^2$, multiplied by an unknown overall
coefficient $\zeta$, write the field redefinition
(\ref{eq:fieldred2}) in terms of the unknown $\gamma$, plug in the
field redefinition, and match with the effective action
(\ref{eq:2der}), which would allow us to fix $\gamma$ and $\zeta =
-1/4$.

A more direct approach can be used when we have an explicit expression
for the gauge invariance of the effective theory.  In this case we can
simply try to construct a field redefinition which relates this
invariance to the usual Yang-Mills gauge invariance.  When finding the
field redefinition relating the deformed and undeformed theories,
however, a further subtlety arises, which was previously encountered
in related situations \cite{Ghoshal-Sen,David-U}.  Namely, there
exists for any theory a class of trivial gauge invariances.  Consider
a theory with fields $\phi_i$ and action $S(\phi_i)$.  This theory has
trivial gauge transformations of the form
\begin{equation}
\label{eq:trivial} \delta \phi_i = \mu_{ij} \frac{\delta S}{\delta
\phi_j}
\end{equation}
where $\mu_{ij} = - \mu_{ji}$.  Indeed, the variation of the
action under this transformation is $\delta S = \mu_{ij}
\frac{\delta S}{\delta \phi_i} \frac{\delta S}{\delta \phi_j} =
0$.  These transformations are called trivial because they do not
correspond to a constraint in the Hamiltonian picture.  The
conserved charges associated with trivial transformations are
identically zero.  In comparing the gauge invariance of the
effective action $S[A]$ to that of the Born-Infeld action, we need
to keep in mind the possibility that the gauge invariances are not
necessarily simply related by a field redefinition, but that the
invariance of the effective theory may include additional terms of
the form (\ref{eq:trivial}).  In considering this possibility, we
can make use of a theorem (theorem 3.1 of
\cite{Henneaux-Teitelboim}), which states that under suitable
regularity assumptions on the functions $\frac{\delta S}{\delta
\phi_i}$ any gauge transformation that vanishes on shell can be
written in the form (\ref{eq:trivial}).  Thus, when identifying
the field redefinition transforming the effective vector field $A$
to the gauge field $\hat{A}$, we allow for the possible addition
of trivial terms.  

The benefit of the first method described above for determining the
field redefinition is that we do not need to know the
explicit form of the gauge transformation.  Once the field
redefinition is known we can find the gauge transformation law in
the effective theory of $A_\mu$ up to trivial terms by plugging
the field redefinition into the standard gauge transformation law
of $\hat A_\mu$.  In the explicit example of $\partial^2A^4$ terms
considered in the previous  subsection
we determined that the gauge transformation of the vector field
$A_\mu$ is given by
\begin{equation}
\delta A_\mu = \partial_\mu \lambda -g_{{\rm YM}}^2 \gamma (A^2
\partial_\mu \lambda -2A_\mu A_\nu \partial^\nu \lambda)
\end{equation}
plus possible trivial terms which we did not consider.  We have found
the numerical value of $\gamma$ in (\ref{eq:gamma}).  If we had been
able to directly compute this gauge transformation law, finding the
field redefinition (\ref{eq:fieldred2}) would have been
trivial.  Unfortunately, as we shall see in a moment, the procedure for
computing the higher-order terms in the gauge invariance of the
effective theory is complicated to implement, which makes the second
method less practical in general for determining the field
redefinition.  We can, however, at least compute the terms in the gauge
invariance which are of order $A \lambda$ directly from the definition
(\ref{eq:gaugetransform}).  Thus, for these terms the second method
just outlined for computing the field redefinition can be used.  We use
this method in section \ref{sec:nonabelianD3A3} to compute the field
redefinition including terms at order $\partial A^2$ and $\partial^2
A$ in the nonabelian theory.

Let us note that the  field redefinition that makes the gauge
transformation standard is not unique.  There is a class of field
redefinitions that preserves the gauge structure and mass-shell
condition
\begin{align}
\label{eq:consredef}
\hat A'_\mu &= \hat A_\mu + T_\mu(\hat F) + \hat{D}_\mu \xi(\hat A)
\nonumber ,  \\
\hat \ld' &= \hat \ld + \delta_{\hat \ld} \xi(\hat A_\mu) \,.
\end{align}
In this field redefinition $T_\mu(\hat F)$ depends on $\hat{A}_\mu$ only
through the covariant field strength and its covariant
derivatives.  The term $\xi$ is a trivial (pure gauge) field
redefinition, which is essentially a gauge transformation with
parameter $\xi(A)$.  The resulting ambiguity in the effective
Lagrangian has a field theory interpretation based on the equivalence
theorem \cite{equivalencetheorem}.  According to this theorem,
different Lagrangians give the same S-matrix elements if they are
related by a change of variables in which both fields have the same
gauge variation and satisfy the same mass-shell condition.

Let us now describe briefly how the different forms of gauge
invariance arise in the world-sheet and string field theory approaches
to computing the vector field action.  
We primarily carry
out this discussion in the context of the abelian theory, although
similar arguments can be made in the nonabelian case.
In a world-sheet sigma model
calculation one introduces the boundary interaction term
\begin{equation}
\oint_\gamma A_\mu \frac{d X^\mu}{d \tau} d \tau.
\end{equation}
This term is explicitly invariant under
\begin{equation}
A_\mu \rightarrow A_\mu+\pd_\mu \ld.
\end{equation}
Provided that one can find a systematic method of calculation that
respects this gauge invariance, the resulting effective action
will possess this gauge invariance as well.  This is the reason
calculations such as those in \cite{Fradkin-Tseytlin85II,acny}
give an effective action with the usual gauge invariance.

In the cubic SFT calculation, on the other hand, the gauge
invariance is much more complicated.  The original theory has an
infinite number of gauge invariances, given by
(\ref{eq:gaugetransform}).  We have fixed all but one of these
gauge symmetries;  the remaining symmetry comes from a gauge
transformation that may change the field $\alpha$, but which keeps
all other auxiliary fields at zero.  A direct construction of this
gauge transformation in the effective theory of $A_\mu$ is rather
complicated, but can be described perturbatively in three steps:
\begin{enumerate}
\item Make an SFT gauge transformation (in the full theory with an
infinite number of fields) with the parameter
\begin{equation}
\label{eq:gaugeparameter} \ket{\Lambda'}=\frac{i}{\sqrt{2}}
\lambda(x) b_{-1}\ket{0}.
\end{equation}
This gauge transformation transforms  $\alpha$ and $A_\mu$ as
\begin{align}
\label{eq:gaugetemplate}
\delta &  A_\mu = \pd_\mu \ld + i g_{YM} ( \cdots), \nonumber \\
\delta & \alpha = \sqrt{2}\, \pd^2 \ld + i g_{YM}( \cdots),
\end{align}
and transforms all fields in the theory in a computable fashion.

\item The gauge transformation $\ket{\Lambda'}$ takes us away from
the gauge slice we have fixed by generating fields associated with
states containing $c_0$ at all higher levels.  We now have to make
a second gauge transformation with a parameter
$\ket{\Lambda''(\ld)}$ that will restore our gauge of choice.  The
order of magnitude of the auxiliary fields we have generated at
higher levels is $ {\cal O} (g \ld \Phi)$.  Therefore
$\ket{\Lambda''(\ld)}$ is of order $g \ld \Phi$.  Since we already
used the gauge parameter at level zero, we will choose
$\ket{\Lambda''}$ to have nonvanishing components only for massive
modes.  Then this gauge transformation does not change the
massless fields linearly, so the contribution to the gauge
transformation at the massless level will be of order ${\cal O}
(g^2 \ld \Phi^2)$.  The gauge transformation generated by
$\ket{\Lambda'' (\lambda)}$ can be computed as a perturbative
expansion in $g$.  Combining this with our original gauge
transformation generated by $\ket{\Lambda'}$ gives us a new gauge
transformation which transforms the massless fields linearly
according to (\ref{eq:gaugetemplate}), but which also keeps us in
our chosen gauge slice.

\item In the third step we eliminate all the fields besides
$A_\mu$ using the classical equations of motion.  The SFT equations
of motion are
\begin{equation}
Q_{B}\ket{\Phi} =-g \langle  \Phi, \Phi \ket{V_3} .
\end{equation}
The BRST operator preserves the level of fields; therefore, the
solutions for massive fields and $\alpha$ in terms of $A_\mu$ will
be of the form
\begin{align}
&\psi_{\mu_1, \ldots,\mu_n} = {\cal O} (g A^{2}), \\
&\alpha = \frac{1}{\sqrt{2}} \pd\cdot A + {\cal O} (g A^2)
\end{align}
where $\psi_{\mu_1, \ldots,\mu_n}$ is a generic massive field.
Using these EOM to eliminate the massive fields and $\alpha$ in
the gauge transformation of $A_\mu$ will give terms of order
${\cal O} (g^2 A^2)$.
\end{enumerate}

To summarize, the gauge transformation in the effective theory for
$A_\mu$ is of the form
\begin{align}
\label{eq:gaugetemplateII}
\delta A_{\mu} =  \pd_\mu \ld + R_\mu(A, \ld), 
\end{align}
where $R_\mu$ is a specific
function of $A$ and $\ld$ at order $g^2 A^2 \lambda$, which can in
principle be computed using the method just described.  In the
nonabelian theory, there will also be terms at order $gA \lambda$
arising directly from the original gauge transformation
$\ket{\Lambda}$; these terms are less complicated and can be computed
directly from the cubic string field vertex.

In this subsection, we have discussed two approaches to computing
the field redefinition which takes us from the effective action
$S[A]$ to a covariant action written in terms of the gauge field
$\hat{A}$, which should have the form of the Born-Infeld action
plus derivative corrections.  In the following sections we use
these two approaches to check that various higher-order terms in
the SFT effective action indeed agree with known terms in the
Born-Infeld action, in both the abelian and nonabelian theories.

\subsection{Terms of the form $\partial^4A^4$}
\label{sec:abelian4der} The goal of this subsection is to verify
that after an appropriate field redefinition the $\pd^4 A^4$ terms
in the abelian effective action derived from SFT transform into
the $ \hat F^4 - \quarter (\hat F^2)^2 $ terms of the Born-Infeld
action (including the correct constant factor of $(2 \pi
g_{YM})^2/8$).  To demonstrate this, we use the first method
discussed in the previous subsection.  Since the total number of
$\pd^4 A^4$ terms is large we restrict attention to a subset of
terms: namely those terms where indices on derivatives are all
contracted together.  These terms are independent from other terms
at the same order in the effective action.
By virtue of the equations of motion (\ref{eq:alphaEOM}) the
diagrams with $\alpha$ do not contribute to these terms.  This
significant simplification is the reason why we choose to
concentrate on these terms.  Although we only compute a subset of
the possible terms in the effective action, however, we find that
these terms are sufficient to fix both coefficients in the
Born-Infeld action at order $F^4$.

The terms we are interested in have the general form
\begin{multline}
\label{eq:p4A4} S^{4}_{(\partial \cdot \partial)A^4}
 = g^2 \int d^{26} x \bigl ( d_1 (\pd_\mu A_\ld \pd^\mu
A^\ld)^2 + d_2 \pd_\mu A_\ld \pd_\nu A^\ld  \pd^\mu A_\sigma
\pd^\nu A^\sigma \\  + d_3 A_\ld \pd_\nu A^\ld \pd_\mu A_\sigma
\pd^\mu \pd^\nu A^\sigma + d_4 \pd_\mu A_\ld
\pd_\nu A^\ld A_\sigma \pd^\mu \pd^\nu A^\sigma \\
+ d_5  A_\ld  A^\ld \pd_\mu \pd_\nu  A_\sigma \pd^\mu \pd^\nu
A^\sigma + d_6 A_\ld \pd_\mu \pd_\nu A^\ld  A_\sigma \pd^\mu
\pd^\nu A^\sigma \bigr).
\end{multline}
The coefficients for these terms in the effective action are given
by
\begin{equation}
d_i = \frac{1}{2}\, \cN^2 \int_0^\infty d\tau e^{\tau} \Det \left
(\frac{1 - \td X^2}{(1- \td V^2)^{13}} \right) P_i^{(4)}(A, B, C)
\end{equation}
with
\begin{align}
P_1^{(4)} & =  P_5^{(4)} =
A_{11}^2 A_{00}^2 + B_{11}^2 B_{00}^2 +  C_{11}^2 C_{00}^2,            \nonumber  \\
P_2^{(4)} & =  P_6^{(4)} = A_{11}^2 \left(B_{00}^2 + C_{00}^2
\right) + B_{11}^2 \left(A_{00}^2 + C_{00}^2 \right) +
C_{11}^2 \left(A_{00}^2 + B_{00}^2 \right), \nonumber  \\
P_3^{(4)} & = 4 A_{11}^2 A_{00} \left (B_{00} + C_{00} \right) + 4
B_{11}^2 B_{00} \left (A_{00} + C_{00} \right) +
4 C_{11}^2 C_{00} \left (A_{00} + B_{00} \right),  \\
P_4^{(4)} & = 4 A_{11}^2 B_{00} C_{00} + 4 B_{11}^2 A_{00} C_{00}
+ 4 C_{11}^2 A_{00} B_{00}.  \nonumber
\end{align}
Computation of the integrals gives us
\begin{align}
d_1 &= d_5 \approx 3.14707539, &  d_3 &\approx 18.51562023, \nonumber  \\
d_2 &= d_6 \approx 2.96365920, &  d_4 &\approx 0.99251621.
\end{align}
To match these coefficients with the BI action we need to write
the general field redefinition to  order $\partial^2 A^3$
(again, keeping only terms with all derivatives contracted)
\begin{multline}
\label{eq:chvar} \hat A_{\mu}= A_{\mu} + g^2 \bigl(\gamma A^{2}
A_{\mu} + \alpha_{1} A_{\mu}A_{\sigma}\pd^{2}A^{\sigma} +\alpha_{2}
A^{2}\pd^{2}A_{\mu}\\ +\alpha_{3}
A_{\mu}\pd_{\ld}A_{\sigma}\pd^{\ld}A^{\sigma} + \alpha_{4}
A_{\sigma}\pd_{\ld}A_{\mu}\pd^{\ld}A^{\sigma} \bigr ).
\end{multline}
Using the general theorem quoted in the previous subsection, we know
that there is a field redefinition relating the action containing the
terms (\ref{eq:p4A4}) to a covariant action written in terms of a
conventional field strength $\hat F$.  The coefficients of $\hat F^2$
and $\hat{F}^3$ are already fixed, so the most generic action up to
$\hat F^4$ is
\begin{equation}
\Tr \int d x \Bigl(-\quarter \hat F^2 + g^2 \left( a \hat F^4 + b
\bigl(\hat F^2\bigr)^2 \right ) + O\bigl(\hat F^6\bigr) \Bigr).
\label{eq:general-f4}
\end{equation}
We plug  the change of variables (\ref{eq:chvar}) into this
equation and collect $\pd^4 A^{4}$  terms with derivatives
contracted together:
\begin{equation}
\begin{split}
\label{eq:p4A4exp} g^2 \int d^{26} x \Bigl ( (\alpha_1 - \alpha_3
& + 4 b) (\pd_\mu A_\ld \pd^\mu A^\ld)^2 \\ & + (\alpha_1 + 2
\alpha_2 - \alpha_4 + 2a) \pd_\mu A_\ld \pd_\nu A^\ld  \pd^\mu
A_\sigma \pd^\nu A^\sigma \\  & + (4 \alpha_1 + 4 \alpha_2 - 2
\alpha_3 - \alpha_4) A_\ld \pd_\nu A^\ld \pd_\mu A_\sigma \pd^\mu
\pd^\nu A^\sigma \\ & \ \ \quad \  + (2 \alpha_1 + 2\alpha_2 -
\alpha_4) \pd_\mu A_\ld \pd_\nu A^\ld A_\sigma \pd^\mu \pd^\nu
A^\sigma \\ & \qquad + \alpha_2  A_\ld  A^\ld \pd_\mu \pd_\nu
A_\sigma \pd^\mu \pd^\nu A^\sigma + \alpha_1 A_\ld \pd_\mu \pd_\nu
A^\ld  A_\sigma \pd^\mu \pd^\nu A^\sigma \Bigr).
\end{split}
\end{equation}
The assumption that (\ref{eq:p4A4}) can be written as
(\ref{eq:p4A4exp}) translates into a system of linear equations
for $a$, $b$ and $\alpha_1, \dots \alpha_4$ with the right hand
side given by $d_1, \dots d_6$.  This system is non-degenerate and
has a unique solution
\begin{align}
\alpha_1 &= d_6 \approx  2.9636592,  \nonumber \\
\alpha_2 &= d_5 \approx  3.1470754,  \nonumber \\
\alpha_3 &= \half (-d_3 + d_4 + 2 d_5 +2 d_6) \approx - 2.6508174, \nonumber \\
\alpha_4 &= - d_4 +2 d_5 +2 d_6 \approx  11.2289530,   \\
a &= \half (d_2 - d_4 + d_6) \approx  2.4674011,  \nonumber \\
b &= \frac{1}{8} (2 d_1 - d_3 + d_4 + 2 d_5) \approx  -
0.6168503.  \nonumber
\end{align}
This determines the coefficients $a$ and $b$ in the effective
action (\ref{eq:general-f4}) to 8 digits of precision.  These
values agree precisely with those that we expect from the
Born-Infeld action, which are given by
\begin{align}
a &= \frac{\pi^2}{4} \approx 2.4674011, \nonumber \\
b &=  -  \frac{\pi^2}{16} \approx - 0.6168502.
\end{align}
Thus, we see that after a field redefinition, the effective vector
theory derived from string field theory agrees with Born-Infeld to
order $F^4$, and correctly fixes the coefficients of both terms at
that order.  This calculation could in principle be continued to
compute higher-derivative corrections to the Born-Infeld action of the
form $\partial^6A^4$ and higher, but we do not pursue such
calculations further here.

Note that, assuming we know that the Born-Infeld action takes the
form
\begin{equation}
S_{BI} = - T \int dx \sqrt{- \det\left(\eta^{\mu\nu} + T^{- \half}
F^{\mu\nu}\right)} .
\end{equation}
with undetermined D-brane tension, we can fix $T = 1 / (2
\pi\alpha' g_{YM})^2$ from the coefficients at $F^2$ and $F^4$.  We
may thus think of the calculations done so far as providing
another way to determine the D-brane tension from SFT.

\subsection{Terms of the form $A^{2n}$}
\label{sec:a-higher} In the preceding discussion we have focused
on terms in the effective action which are at most quartic in the
vector field $A_\mu$.  It is clearly of interest to extend this
discussion to terms of higher order in $A$.  A complete analysis
of higher-order terms, including all momentum dependence, involves
considerable additional computation.  We have initiated analysis
of higher-order terms by considering the simplest class of such
terms: those with no momentum dependence.  As for the quartic
terms of the form $(A^\mu A_{\mu})^2$ discussed in Section 4.2, we
expect that all terms in the effective action of the form
\begin{equation}
(A^\mu A_{\mu})^{n} \label{eq:a2n}
\end{equation}
should vanish identically when all diagrams are considered.  In
this subsection we consider terms of the form (\ref{eq:a2n}).  We
find good numerical evidence that these terms indeed vanish, up to
terms of the form $A^{10}$.

In Section 4.2 we found strong numerical evidence that the term
(\ref{eq:a2n}) vanishes for $n = 2$ by showing that the coefficient
$\gamma_+$ in (\ref{eq:quarticterm}) approaches 0 in the
level-truncation approximation.  This $A^4$ term involves only one
possible diagram.  As $n$ increases, the number of diagrams involved
in computing $A^{2n}$ increases exponentially, and the complexity of
each diagram also increases, so that the primary method used in this
paper becomes difficult to implement.  To study the terms
(\ref{eq:a2n}) we have used a somewhat different method, in which we
directly truncate the theory by only including fields up to a fixed
total oscillator level, and then computing the cubic terms for each of
the fields below the desired level.  This was the original method of
level truncation used in \cite{Kostelecky-Samuel} to compute the
tachyon 4-point function, and in later work \cite{Sen-Zwiebach,
Moeller-Taylor} on level truncation on the problem of tachyon
condensation.  As discussed in Section 3.3, the method we are using
for explicitly calculating the quartic terms in the action involves
truncating on the level of the intermediate state in the 4-point
diagram, so that the two methods give the same answers.  While level
truncation on oscillators is very efficient for computing low-order
diagrams at high level, however, level truncation on fields is more
efficient for computing high-order diagrams at low level.

In \cite{Moeller-Taylor}, a recursive approach was used to
calculate coefficients of $\phi^n$ in the effective tachyon
potential from string field theory using level truncation on
fields.  Given a cubic potential
\begin{equation}
V = \sum_{i, j}d_{ij} \,\psi_{i} \psi_j +  \sum_{i, j, k} g t_{ijk}\,
\psi_i \psi_j \psi_k
\end{equation}
for a finite number of fields $\psi_i, i = 1, \ldots, N$ at $p =
0$, the effective action for $a =\psi_1$ when all other fields are
integrated out is given by
\begin{equation}
V_{{\rm eff}} (a) = \sum_{n = 2}^{ \infty}  \frac{1}{n}
v^{1}_{n-1} a^n g^n
\end{equation}
where $v^{i}_{n}$ represents the summation over all graphs with
$n$ external $a$ edges and a single external $\psi^i$, with no
internal $a$'s.  The $v's$ satisfy the recursion relations
\begin{eqnarray}
v_1^i & = &  \delta^i_1 \nonumber\\
v_n^i & = & \frac{3}{2} \sum_{m = 1}^{n-1}  d^{ij}\, t_{jkl}\,
\hat{v}_m^k \hat{v}_{n-m}^l \label{eq:recursion}
\end{eqnarray}
where $d^{ij}$ is the inverse matrix to $d_{ij}$ and
\begin{equation}
\hat{v}_n^i = \left\{
\begin{array}{l}
0, \;\;\;\;\;i = 1 \; {\rm and} \; n > 1\\
v_n^i, \,\;\;\; {\rm otherwise}
\end{array}
\right.
\end{equation}
has been defined to project out internal $a$ edges.

We have used these relations to compute the effective action for
$A_\mu$ at $p = 0$.  We computed all quadratic and cubic
interactions between fields up to level 8 associated with states
which are scalars in 25 of the space-time dimensions and which
include an arbitrary number of matter oscillators $a^{25}_{-n}$.
Plugging the resulting quadratic and cubic coefficients into the
recursion relations (\ref{eq:recursion}) allows us to compute the
coefficients $c_{2n} = v^{1}_{2n-1}/2n$ in the effective action
for the gauge field $A_\mu$
\begin{equation}
\sum_{n = 1}^{ \infty}  -c_{2n} g^n (A^{\mu} A_{\mu})^{n}
\end{equation}
for small values of $n$ .  We have computed these coefficients up
to $n =7$ at different levels of field truncation up to $L = 8$.
The results of this computation are given in
Table~\ref{t:higher-order} up to $n = 5$, including the predicted
value at $L = \infty$ from a $1/L$ fit to the data at levels 2, 4,
6 and 8.
\begin{table}[htp]
\begin{center}
\begin{tabular} {|l|l | l | l | l|}
\hline \hline
Level & $c_4$ &   $c_6$ &  $c_8$ & $c_{10}$ \\
\hline \hline
2  &  0.200  &  1.883  & 6.954  & 28.65 \\
\hline
4  &  0.097  &  1.029  & 6.542 & 37.49 \\
\hline
6  &  0.063  &  0.689  &  5.287  & 37.62 \\
\hline
8  &  0.046  &  0.517  &  4.325  & 34.18 \\
\hline \hline
$\infty$  &  0.001  &  0.014  &  -0.229 & 1.959 \\
\hline \hline
\end{tabular}
\end{center}
\caption{\footnotesize Coefficients of $A^{2n}$ at various levels
of truncation} \label{t:higher-order}
\end{table}
The results in Table~\ref{t:higher-order} indicate that, as
expected, all coefficients $c_{2n}$ will vanish when the level is
taken to infinity.  The initial contribution at level 2 is
canceled to within $0.6\%$ for terms $A^{4}$, within 0.8\% for
terms $A^{6}$, within 4\% for terms $A^8$, and within 7\% for
terms $A^{10}$.  It is an impressive success of the
level-truncation method that for $c_{10}$, the cancellation
predicted by the $1/L$ expansion is so good, given that the
coefficients computed in level truncation {\it increase} until
level $L = 8$.  We have also computed the coefficients for larger
values of $n$, but for $n > 5$ the numerics are less compelling.
Indeed, the approximations to the coefficients $c_{12}$ and beyond
continue to grow up to level 8.  We expect that a good prediction
of the cancellation of these higher-order terms would require
going to higher level.

The results found here indicate that the method of level
truncation in string field theory seems robust enough to correctly
compute higher-order terms in the vector field effective action.
Computing terms with derivatives at order $A^6$ and beyond would
require some additional work, but it seems that a reasonably
efficient computer program should be able to do quite well at
computing these terms, even to fairly high powers of $A$.

\section{The nonabelian Born-Infeld action}
\label{sec:nonabelianBI}

We now consider the theory with a nonabelian gauge group.  As we
discussed in section \ref{sec:Born-Infeld}, the first term beyond
the Yang-Mills action in the nonabelian analogue of the
Born-Infeld action has the form ${\rm Tr}\; \hat{F}^3$.  As in the
previous section, we expect that a field redefinition is necessary
to get this term from the effective nonabelian vector field theory
derived from SFT.  In this section we compute the terms in the
effective vector field theory to orders $\partial^3A^3$ and
$\partial^2 A^4$, and we verify that after a field redefinition
these terms reproduce the corresponding pieces of the $\hat{F}^3$
term, with the correct coefficients.  In section
\ref{sec:nonabelianD3A3} we analyze $\pd^3 A^3$ terms, and in
subsection \ref{sec:nonabelianD2A4} we consider the $\pd^2 A^4$
terms.

\subsection{$\pd^3  A^3$ terms}
\label{sec:nonabelianD3A3}

In section \ref{sec:YMcubic} we showed that the terms of the form
$\partial A^3$ in the nonabelian SFT effective action for $A$
contribute to the $\hat{F}^2$ term after a field redefinition.  We
now consider terms at order $\partial^3 A^3$.  Recall from
(\ref{eq:cubicmassless}) and (\ref{eq:cubicAa2}) that the full
effective action for $\alpha$ and $A$ at this order is given by
\begin{multline}
\check{S}_{A^3}[A, \alpha] =
i g_{YM} \int d x \Tr\Bigl(\frac{1}{6} \bigl (\pd_\lambda \td
A^\mu \pd_\mu \td A^\nu \pd_\nu \td A^\lambda - \pd_\nu \td A^\mu
\pd_\lambda \td A^\nu \pd_\mu \td A^\lambda \bigr)  \\
 - \pd_\mu \td A_\nu [\td A^\mu, \td A^\nu]
+ \half [\td A_\nu, \pd^\ld \td A_\mu] \pd^{\mu} \pd^{\nu} \td
A_\ld + \td A^\mu \left[\pd_\mu\td\alpha ; \td \alpha \right]
 \Bigr )\label{eq:cubic}
\end{multline}
where $\td A_\mu = \exp(-\half V^{11}_{00} \pd^2) A_\mu$, and
similarly for $\tilde{\alpha}$.  After eliminating $\alpha$ using
(\ref{eq:quadraticmassless}) and (\ref{eq:cubic}) and
integrating by parts to remove terms containing $\pd A $, we find
that the complete set of terms at order $\partial^3A^3$ is given
by
\begin{multline}
\label{eq:3D3GB} 
S^{[3]}[A]_{A^3}=
i g_{YM} \int d x \Tr\Bigl(\frac{2}{3}\bigl(
\pd_\lambda  A^\mu \pd_\mu  A^\nu \pd_\nu  A^\lambda - \pd_\nu
A^\mu \pd_\lambda  A^\nu \pd_\mu
A^\lambda\bigr)\\
+\frac{1}{2} V^{11}_{00}\bigl(
 \pd_\mu \partial^2 A_\nu [ A^\mu,  A^\nu] +
 \pd_\mu A_\nu [\partial^2 A^\mu,  A^\nu] +
 \pd_\mu A_\nu [ A^\mu,  \partial^2 A^\nu]
\bigr) \Bigr).
\end{multline}
Note that unlike the quartic terms in $A$, our expressions for
these terms are exact.

Let us now consider the possible terms that we can get after the
field redefinition to the field $\hat{A}$ with standard gauge
transformation rules.  Following the analysis of
\cite{Tseytlin86}, we write the most general covariant action to
order $\hat F^3$ (keeping $D$ at order $F^{1/2}$ as discussed
above)
\begin{equation}
\label{eq:3dercorr} -\frac{1}{4}\hat{F}^2+ig_{YM}a\hat{F}^{3}+\chi
\hat{D}_{\sigma}\hat{F}^{\sigma\mu} \hat{D}^{\nu}\hat{F}_{\nu\mu}+{\cal
O}(\hat{F}^{4}),
\end{equation}
where
\begin{equation}
\hat{D}_{\mu}=\pd_{\mu}-ig_{YM}[\hat A_{\mu},\cdot\  ].
\end{equation}
The action (\ref{eq:3dercorr}) is not invariant under field
redefinitions which keep the gauge invariance unchanged.  Under
the field redefinition
\begin{equation}
\label{eq:consredefexplicit} \hat{A}_{\mu}'= \hat{A}_{\mu}+\upsilon
\hat{D}_{\sigma}\hat{F}^{\sigma}_{\,\mu}.
\end{equation}
we have
\begin{align}
a' &= a, \nonumber \\
\chi' & = \chi - \upsilon.
\end{align}
Thus, the coefficient $a$ is  defined unambiguously, while $\chi$
can be set to any chosen value by a field redefinition.
Just as we have an exact formula for the cubic terms in the SFT
action, we can also compute the gauge transformation rule exactly
to quadratic order in $A$ using (\ref{eq:gaugetransform}).
After some calculation, we find that the gauge variation for
$A_\mu$ to order $A^2 \ld$ is given by (before integrating out
$\alpha$)
\begin{multline}
\delta A_{\mu}  = \pd_{\mu}\ld - i g_{YM} \Bigl([A_\mu, \ld]_\star
- [\pd_\mu A_\nu, \pd^\nu \ld]_\star + [ A^\nu, \pd_\mu \pd_\nu
\ld]_\star + \\ \frac{1}{\sqrt{2}} [\pd_\mu  B,  \ld]_\star -
\frac{1}{\sqrt{2}} [ B,  \pd_\mu  \ld]_\star \Bigr).
\end{multline}
where $B = \alpha - \frac{1} {\sqrt{2}} \pd_ \mu A^\mu$ as in
section (\ref{sec:YMquadratic}).  The commutators are taken with
respect to the product
\begin{equation}
f(x) \star g(x) = f(x) e^{- V^{11}_{00}( \overleftarrow{\pd}^2 +
\overleftarrow{\pd} \cdot \overrightarrow{\pd}
+\overrightarrow{\pd}^2)} g(x).
 \end{equation}
The equation of motion for $\alpha$ at leading order is simply $B
= 0$.  Eliminating $\alpha$ we therefore have
\begin{equation}
\label{eq:gggtt} \delta A_\mu  = \pd_{\mu}  \ld - i g_{YM} \Bigl([
A_\mu,  \ld]_\star + [\pd^\nu  \ld, \pd_\mu  A_\nu]_\star + [
A^\nu, \pd_\mu \pd_\nu  \ld]_\star  \Bigr).
\end{equation}
We are interested in considering the terms at order $\partial^2 A
\lambda$ in this gauge variation.  Recall that in section
\ref{sec:fieldredefinitions} we observed that the gauge
transformation may include extra trivial terms which vanish on shell.
Since the leading term in the equation of motion for $A$ arises at
order $\partial^2 A$, it is possible that (\ref{eq:gggtt}) may
contain a term of the form
\begin{equation}
\label{eq:gttrivexpl} \delta A_\mu =  \rho\, [\ld, \pd^2 A_\mu -
\pd_\mu \pd \cdot A] + \cO(\ld A^2)
\end{equation}
in addition to a part which can be transformed into the standard
nonabelian gauge variation through a field redefinition.  Thus, we
wish to consider the one-parameter family of gauge transformations
%
%
%
\begin{multline}
\label{eq:gtexpanded} \delta  A = \pd_{\mu} \ld - i g_{YM}
\bigl([{ A}_\mu,  \ld] - V^{11}_{00} [\pd^2 { A}_\mu,  \ld] \\ -
V^{11}_{00} [\pd_\nu { A}_\mu,  \pd^\nu \ld] - V^{11}_{00} [
A_\mu,  \pd^2 \ld] + \rho\, [\ld, \pd^2  A_\mu - \pd_\mu \pd \cdot
A] + {\cal O}(\ld  A^2, \lambda \partial^4A) \bigr ),
\end{multline}
where $\rho$ is an as-yet undetermined constant.  We now need to
show, following the second method discussed in subsection
\ref{sec:fieldredefinitions}, that there exists a field
redefinition which takes a field $A$ with action (\ref{eq:3D3GB})
and a gauge transformation of the form (\ref{eq:gtexpanded}) to a
gauge field $\hat{A}$ with an action of the form
(\ref{eq:3dercorr}) and the standard nonabelian gauge
transformation rule.

The leading terms in the field redefinition can be parameterized
as
\begin{align}
\label{eq:redefnonabelian}  \hat{A}_{\mu} &= A_{\mu} +
\upsilon_{1}\pd_{\mu}\pd\cdot A + \upsilon_{2}\pd^{2}
A_{\mu}+ig_{YM}\big(\upsilon_{3} [A_{\sigma},\pd_{\mu} A^{\sigma}]
+ \upsilon_{4} [A_{\mu},\pd\cdot A]+\upsilon_{5}[\pd_{\sigma}
A_{\mu},
A^{\sigma}]\big),\nonumber \\
 \hat{\ld} &=\ld+\upsilon_{6}\pd^{2}\ld + ig_{YM}\big(\upsilon_{7}
[\pd\cdot  A,\ld]+\upsilon_{8}[A_{\sigma},\pd^{\sigma}\ld]\big).
\end{align}
The coefficient $\upsilon_1$ can be chosen arbitrarily through a
gauge transformation, so we simply choose $\upsilon_1 =
-\upsilon_2$.  The requirement that the RHS of
(\ref{eq:redefnonabelian}) varied with (\ref{eq:gtexpanded}) and
rewritten in terms of $\hat A$, $\hat \ld$ gives the standard
transformation law for $\hat A$, $\hat \ld$ up to terms of order
$\cO(\hat \ld \hat A^2)$ gives a system of linear equations with
solutions depending on one free parameter $\upsilon$.
\begin{align}
\upsilon_2 &= -\upsilon_1 = \upsilon, & \rho &= V_{00}^{11}, \nonumber \\
\upsilon_3 & = 1 - \half V_{00}^{11} + \upsilon, &  \upsilon_6 & = 0, \nonumber            \\
\upsilon_4 & = - V_{00}^{11} + \upsilon,    &     \upsilon_7 & = V_{00}^{11},    \\
\upsilon_5 & = - V_{00}^{11}+2 \upsilon,   &     \upsilon_8 & =
\half V_{00}^{11}\,. \nonumber
\end{align}

It is easy to see that the parameter $\upsilon$ generates the
field redefinition (\ref{eq:consredefexplicit}).
For simplicity, we set $\upsilon = 0$.
The field redefinition is then given by
\begin{equation}
\label{eq:frdf} \hat{A}_{\mu}=  A_{\mu}  - ig_{YM}
\Bigl(\bigr(\half  V^{11}_{00} - 1 \bigl)[A_{\sigma}, \pd_{\mu}
A^{\sigma}] +
 V^{11}_{00} [A_{\mu}, \pd \cdot A] +
V^{11}_{00} [\pd_{\sigma} A_{\mu}, A^{\sigma}]\Bigr) \,.
\end{equation}
We can now  plug in  the field redefinition (\ref{eq:frdf}) into
the action (\ref{eq:3dercorr}) and compare with the
$\partial^3A^3$ term in the SFT effective action (\ref{eq:3D3GB}).
We find agreement when the coefficients in (\ref{eq:3dercorr}) are
given by
\begin{equation}
a = \frac{2}{3}, \;\;\;\;\; \chi = 0.
\end{equation}
Thus, we have shown that the terms of order $\partial^3A^3$ in the
effective nonabelian vector field action derived from SFT are in
complete agreement with the first nontrivial term in the nonabelian
analogue of the Born-Infeld theory, including the overall constant.
Note that while the coefficient of $a$ agrees with that in
(\ref{eq:BI+DC}), the condition $\chi = 0$ followed directly from our
choice $\upsilon = 0$; other choices of $\upsilon$ would lead to other
values of $\chi$, which would be equivalent under the field
redefinition (\ref{eq:consredefexplicit}).

\subsection{$\pd^2  A^4$ terms}
\label{sec:nonabelianD2A4} In the abelian theory, the $\pd^2 A^4$
terms disappear after the field redefinition.  In the nonabelian
case, however, the term proportional to $\hat F^3$ contains terms
of the form $\pd^2 {\hat A}^4$.  In this subsection, we show that
these terms are correctly reproduced by string field theory after
the appropriate field redefinition.  Just as in section
\ref{sec:abelian4der}, for simplicity we shall concentrate on the
$\pd^2 A^4$ terms where the Lorentz indices on derivatives are
contracted together.

The terms of interest in the effective nonabelian vector field
action can be written in the form
\begin{multline}
\label{eq:nonabelian2d4Ageneric} S_{A^4}^{[2]}=g_{YM}^2\int
d^{26}x
\biggl(f_{1}\pd_{\sigma}A_{\mu}A^{\mu}\pd^{\sigma}A_{\nu}A^{\nu} +
f_{2} \pd_{\sigma}A_{\mu}A^{\mu}A^{\nu}\pd^{\sigma}A_{\nu}+ f_{3}
A^{\mu}\pd_{\sigma}A_{\mu}A_{\nu}\pd^{\sigma}A^{\nu} \\ + f_{4}
\pd_{\sigma}A_{\mu}\pd^{\sigma}A^{\mu}A_{\nu}A^{\nu}+f_{5}
\pd_{\sigma}A_{\mu}\pd^{\sigma}A_{\nu}A^{\mu}A^{\nu} +
f_{6}\pd_{\sigma}A_{\mu}A^{\nu}\pd^{\sigma}A_{\mu}A^{\nu}\biggl)
\end{multline}
where the coefficients $f_i$ will be determined below.  The
coefficients of the terms in the field redefinition which are
linear and quadratic in $A$ were fixed in the previous subsection.
The relevant terms in the field redefinition for computing  the
terms we are interested in  here are generic terms of order $A^3$
with no derivatives, as well as those from (\ref{eq:frdf}) that do
not have $\pd_\mu$'s contracted with $A_\mu$'s.  Keeping only
these terms we can parametrize the field redefinition as
\begin{equation}
\hat{A}_{\mu}= A_{\mu} +ig_{YM}
(1-\frac{V_{00}^{11}}{2})[A_{\sigma},\pd_{\mu}A^{\sigma}] +
g_{YM}^2 \big(\rho_{1}A_{\sigma}A_{\mu}A^{\sigma}+
\rho_{2}A^{2}A_{\mu}+ \rho_{3}A_{\mu}A^{2}\big).
\end{equation}
In the abelian case this formula reduces to (\ref{eq:fieldred2})
with $\rho_{1}+\rho_{2}+\rho_{3}= 2 \gamma$.  Plugging this field
redefinition into the action
\begin{equation}
\label{eq:actn} \hat S[\hat A_\mu] = \int \Tr\left( - \quarter
\hat F^2 + \frac{2 i}{3} g_{YM} \hat F^3 + {\cal O}({\hat
F}^4)\right).
\end{equation}
and collecting $\pd^2 A^4$ terms with indices on derivatives
contracted together we get

\begin{multline}
\label{eq:nonabelian2d4A} g_{YM}^2\int dx\Bigl[
 (\half V_{00}^{11} - 1 - \rho_{3})
\pd_{\sigma} A_{\mu}A^{\mu}\pd^{\sigma} A_{\nu}A^{\nu} - (\rho_{2}
+ \rho_{3} + V_{00}^{11})
\pd_{\sigma} A_{\mu}A^{\mu}A_{\nu} \pd^{\sigma} A^{\nu} \\
+ (\half V_{00}^{11} - 1 - \rho_{2}) A_{\mu}\pd^{\sigma}
A^{\mu}A_{\nu} \pd^{\sigma} A^{\nu} -
(\rho_{2} + \rho_{3}) \pd_{\sigma} A_{\mu}\pd^{\sigma} A^{\mu}A_{\nu}A^{\nu} \\
+ (2- 2 \rho_{1})\pd_{\sigma} A_{\mu}\pd^{\sigma}
A_{\nu}A^{\mu}A^{\nu} - \rho_{1} \pd_{\sigma}
A_{\mu}A_{\nu}\pd^{\sigma} A^{\mu}A_{\nu}\Bigl].
\end{multline}
Comparing (\ref{eq:nonabelian2d4A}) and
(\ref{eq:nonabelian2d4Ageneric}) we can write the unknown
coefficients in the field redefinition in terms of the $f_i$'s
through
\begin{align}
\label{eq:firicoeff} \rho_{1}&= - f_{6}, & \rho_{2} &= \rho_3 = -
\half f_{4} \,.
\end{align}
We also find a set of constraints on the $f_i$'s which we expect
the values computed from the SFT calculation to satisfy, namely
\begin{align}
\label{eq:frel} f_{1} - \half f_4   &= -1 + \half V_{00}^{11}, &
f_{2} - f_4  & =- V_{00}^{11}, & f_{5}- 2 f_6 &= 2.
\end{align}
On the string field theory side the coefficients $f_i$ are given
by
\begin{equation}
f_{i} =  \frac{1}{2}\, \cN^2 \int_0^\infty d\tau e^{\tau} \Det
\left (\frac{1 - \td X^2}{(1- \td V^2)^{13}} \right) P_{\pd^2 A^4,
i}(A, B, C)
\end{equation}
where, in complete analogy with the previous examples, the
polynomials $P_{\pd^2 A^4, i}$  derived from (\ref{eq:4GB}) and
(\ref{eq:4GBsuppl}) have the form
\begin{align}
P_{\pd^2 A^4, 1} & = -2 \bigl (A_{11}^2 B_{00} + C_{11}^2
B_{00}\bigr), &
P_{\pd^2 A^4, 4} & = -4 \bigl (A_{11}^2 A_{00}+C_{11}^2C_{00}\bigr), \nonumber\\
P_{\pd^2 A^4, 2}& = -4 \bigl ( A_{11}^2 C_{00} + C_{11}^2A_{00}
\bigr), &
P_{\pd^2 A^4, 5} & =  -4B_{11}^2 \bigl ( A_{00} +C_{00}\bigr),                \\
P_{\pd^2 A^4, 3} & =    - 2\bigl(A_{11}^2B_{00}+ C_{11}^2
B_{00}\bigr), & P_{\pd^2 A^4, 6} & =  -4 B_{11}^2B_{00}.  \nonumber
\end{align}
Numerical computation of the integrals gives
\begin{align}
f_{1} &\approx -2.2827697,   & f_{4} \approx  -2.0422916, \nonumber\\
f_{2} &\approx -1.5190433,   & f_{5}  \approx -2.5206270,  \\
f_{3} &\approx -2.2827697,   & f_{6} \approx  -2.2603135.
\nonumber
\end{align}
As one can easily check, the relations (\ref{eq:frel}) are
satisfied with high accuracy.  This verifies that the
$\partial^2 A^4$ terms we have computed in the effective vector
field action are in agreement with the $\hat{F}^3$ term in the
nonabelian analogue of the Born-Infeld action.

\section{Conclusions}
In this paper we have computed the effective action for the
massless open string vector field by integrating out all
other fields in Witten's cubic open bosonic string field theory.  We
have calculated the leading terms in the off-shell action $S[A]$
for the massless vector field $A_\mu$, which we have transformed
using a field redefinition into an action $ \hat{S}[ \hat{ A}]$
for a gauge field $\hat{A}$ which transforms under the standard
gauge transformation rules.  For the abelian theory, we have shown
that the resulting action agrees with the Born-Infeld action to
order $\hat{F}^4$, and that zero-momentum terms vanish to order
$A^{10}$.  For the nonabelian theory, we have shown agreement with
the nonabelian effective vector field action previously computed by
world-sheet methods to order $\hat{F}^3$.
These results demonstrate that string field theory provides a
systematic approach to computing the effective action for massless
string fields.  In principle, the calculation in this paper could
be continued to determine higher-derivative corrections to the
abelian Born-Infeld action and higher-order terms in the
nonabelian theory.

As we have seen in this paper, comparing the string
field theory effective action to the effective gauge theory action
computed using world-sheet methods is complicated by the fact that the
fields defined in SFT are related through a nontrivial field
redefinition to the fields defined through world-sheet methods.  In
particular, the massless vector field in SFT has a nonstandard gauge
invariance, which is only related to the usual Yang-Mills gauge
invariance through a complicated field redefinition.  This is a
similar situation to that encountered in noncommutative gauge
theories, where the gauge field in the noncommutative theory---whose
gauge transformation rule is nonstandard and involves the
noncommutative star product---is related to a gauge field with
conventional transformation rules through the Seiberg-Witten map.  In
the case of noncommutative Yang-Mills theories, the structure of the
field redefinition is closely related to the structure of the
gauge-invariant observables of the theory, which in that case are
given by open Wilson lines \cite{Liu}.  A related construction
recently appeared in \cite{van-Raamsdonk}, where a field redefinition
was used to construct matrix objects transforming naturally under the
D4-brane gauge field in a matrix theory of D0-branes and D4-branes.
An important outstanding problem in string field theory is to attain a
better understanding of the observables of the theory (some progress
in this direction was made in
\cite{Hashimoto-Itzhaki,Gaiotto-Rastelli-Sen-Zwiebach}).  It seems
likely that the problem of finding the field redefinition between SFT
and world-sheet fields is related to the problem of understanding the
proper observables for open string field theory.

While we have focused in this paper on calculations in the bosonic
theory, it would be even more interesting to carry out analogous
calculations in the supersymmetric theory.  There
are currently several candidates for an open superstring field theory,
including the Berkovits approach \cite{Berkovits} and the (modified)
cubic Witten approach
\cite{Witten:1986qs,Preitschopf:1989gp,Arefeva:1989cm}.  (See
\cite{Ohmori} for further references and a comparison of these
approaches.)  In the abelian case, a superstring calculation should
again reproduce the Born-Infeld action, including all
higher-derivative terms.  In the nonabelian case, it should be
possible to compute all the terms in the nonabelian effective action.
Much recent work has focused on this nonabelian action, and at this
point the action is constrained up to order $F^6$ \cite{k-Sevrin}.  It
would be very interesting if some systematic insight into the form of
this action could be gained from SFT.

The analysis in this paper also has an interesting analogue in the
closed string context.  Just as the Yang-Mills theory describing a
massless gauge field can be extended to a full stringy effective
action involving the Born-Infeld action plus derivative corrections,
in the closed string context the Einstein theory of gravity becomes
extended to a stringy effective action containing higher order terms
in the curvature.  Some terms in this action have been computed, but
they are not yet understood in the same systematic sense as the
abelian Born-Infeld theory.  A tree-level computation in closed string
field theory would give an effective action for the multiplet of
massless closed string fields, which should in principle be mapped by
a field redefinition to the Einstein action plus higher-curvature
terms \cite{Ghoshal-Sen}.  Lessons learned about the nonlocal
structure of the effective vector field theory discussed in this paper
may have interesting generalizations to these nonlocal extensions of
standard gravity theories.

Another direction in which it would be interesting to extend this work
is to carry out an explicit computation of the effective action for
the tachyon in an unstable brane background, or for the combined
tachyon-vector field effective action.  Some progress on the latter
problem was made in \cite{David-U}.  Because the mass-shell condition
for the tachyon is $p^2 = 1$, it does not seem to make any sense to
consider an effective action for the tachyon field, analogous to the
Born-Infeld action, where terms of higher order in $p$ are dropped.
Indeed, it can be shown that when higher-derivative terms are dropped,
any two actions for the tachyon which keep only terms $\partial^k
\phi^{m + k}, m \geq 0,$ can be made perturbatively
equivalent under a field
redefinition (which may, however, have a finite radius of convergence
in $p$).  Nonetheless, a proposal for an effective tachyon + vector
field action of
the form
\begin{equation}
S = V (\phi) \sqrt{-\det (\eta_{\mu \mu} + F_{\mu \nu} + \partial_\mu
  \phi \partial_\nu \phi)} 
\end{equation}
was given in \cite{Garousi,bddep,Kluson} (see also \cite{Sen-action}).
Quite a bit of recent work has focused on this form of effective
action (see \cite{Kutasov-Niarchos} for a recent summary with further
references), and there seem to be many special properties for this
action with particular forms of the potential function $V (\phi)$.  It
would be very interesting to explicitly construct the tachyon-vector
action using the methods of this paper.  A particularly compelling
question related to this action is that of closed string radiation
during the tachyon decay process.  In order to understand this
radiation process, it is necessary to understand back-reaction on the
decaying D-brane \cite{llm}, which in the open string limit
corresponds to the computation of loop diagrams.  Recent work
\cite{Ellwood-Shelton-Taylor} indicates that for the superstring, SFT
loop diagrams on an unstable D$p$-brane with $p < 7$ should be finite,
so that it should be possible to include loop corrections in the
effective tachyon action in such a theory.  The resulting effective
theory should shed light on the question of closed string radiation
from a decaying D-brane.

Ultimately, however, it seems that the most important questions which
may be addressed using the type of effective field theory computed in
this paper have to do with the nonlocal nature of string theory.  The
full effective action for the massless fields on a D-brane, given by
the Born-Infeld action plus derivative corrections, or by the
nonabelian vector theory on multiple D-branes, has a highly nonlocal
structure.  Such nonlocal actions are very difficult to make sense of from the
point of view of conventional quantum field theory.  Nonetheless,
there is important structure hidden in the nonlocality of open string
theory.  For example, the instability associated with contact
interactions between two parts of a D-brane world-volume which are
separated on the D-brane but coincident in space-time is very
difficult to understand from the point of view of the nonlocal theory
on the D-brane, but is implicitly contained in the classical nonlocal
D-brane action.  At a more abstract level, we expect that in any truly
background-independent description of quantum gravity, space-time
geometry and topology will be an emergent phenomenon, not manifest in
any fundamental formulation of the theory.  A nongeometric
formulation of the theory is probably necessary for addressing
questions of cosmology and for understanding very early universe
physics before the Planck time.  It seems very important to develop
new tools for grappling with such issues, and it may be that string
field theory may play an important role in developments in this
direction.  In particular, the way in which conventional gauge theory
and the nonlocal structure of the D-brane action is encoded in the
less geometric variables of open string field theory may serve as a
useful analogue for theories in which space-time geometry and topology
emerge from a nongeometric underlying theory.

\acknowledgments 

We would like to thank Ian Ellwood, Gianluca Grignani, Hong Liu,
Shiraz Minwalla, Martin Schnabl, Ashoke Sen, Jessie Shelton, Mark van
Raamsdonk, Edward Witten and Barton Zwiebach for useful discussions.
This work was supported by the DOE through contract
$\#$DE-FC02-94ER40818.

\appendix

\section{Neumann Coefficients}
In this Appendix we give explicit expressions for and properties of the
Neumann coefficients that we use throughout this paper.  First we
define coefficients $A_n$ and $B_n$ by the series expansions
\begin{align}
\bigg(\frac{1 + i z}{1 - i z}\bigg)^{1/3} &=
\sum_{n\, {\rm even}} A_n z^n + i \sum_{n\, {\rm odd}} A_n z^n, \\
\bigg(\frac{1 + i z}{1 - i z}\bigg)^{2/3} & = \sum_{n\, {\rm even}} B_n
z^n + i \sum_{n\, {\rm odd}} B_n z^n\, .
\end{align}
In terms of $A_n$ and $B_n$ we define the coefficients $N^{r, \pm
s}_{mn}$ as follows:
\begin{align}
N^{r, \pm r}_{nm} & =  \frac{1}{3 (n \pm m)}
 \begin{cases}
 (-1)^n (A_n B_m \pm B_n A_m) & \quad
  m+n \in 2 \bZ ,\ \  m\ne n \\
  0 & \quad  m+n \in 2\bZ +1
 \end{cases}, \nonumber \\
N^{r, \pm (r+1)}_{nm} & =  \frac{1}{6 (n \pm m)}
 \begin{cases}
  (-1)^{n+1} \, (A_n B_m \pm B_n A_m) &
  m+n \in 2 \bZ ,\ \  m\ne n \\
  \sqrt{3} \, (A_n B_m \mp B_n A_m) &
  m+n \in 2\bZ +1
 \end{cases}, \\
N^{r, \pm (r-1)}_{nm} & =  \frac{1}{6 (n \mp m)}
 \begin{cases}
  (-1)^{n+1} (A_n B_m \mp B_n A_m)  &
  m+n \in 2 \bZ ,\ \  m\ne n \nonumber\\
  - \sqrt{3}\, (A_n B_m \pm B_n A_m) &
  m+n \in 2\bZ +1
 \end{cases}\ .
\label{eq:matN}
\end{align}
The coefficients $V^{rs}_{mn}$ are then given by
\begin{subequations}
\begin{align}
&V^{rs}_{nm} = \sqrt{mn}\, \left (N^{r,s}_{nm}+ N^{r,
-s}_{nm}\right) & &
m\ne n,\, m, n > 0\, ,  \\
&V^{rr}_{nn} = \frac{1}{3} \Bigl( 2\sum_{k=0}^n (-1)^{n-k} A_k^2 -
(-1)^n - A_n^2 \Bigr),   &&
n\ne 0\, ,  \\
&V^{r (r+1)}_{nn} = V^{r (r+2)}_{nn} = - \half \bigl((-1)^n +
V^{rr}_{nn}\bigr)  &&   n\ne 0\, , \\
&V^{rs}_{0 n}  = \sqrt{2 n}\, \left (N^{r,s}_{0 n}+ N^{r, -s}_{0
n}\right )
&&  n \ne 0\, ,  \\
&V^{rr}_{00} = - \ln(27/16) \, .
\end{align}
\end{subequations}
The analogous expressions for the ghost Neumann coefficients are
\begin{align}
\cN^{r, \pm r}_{nm} & =  \frac{1}{3 (n \pm m)}
 \begin{cases}
 (-1)^{n+1} (B_n A_m \pm A_n B_m) &
  \ m+n \in 2 \bZ ,\ \  m\ne n \\
  0 & \   m+n \in 2\bZ +1
 \end{cases}, \nonumber \\
\label{eq:ghN} \cN^{r, \pm (r+1)}_{nm} & =  \frac{1}{6 (n \pm m)}
 \begin{cases}
  (-1)^{n} \, (B_n A_m \pm A_n B_m) & \quad
  m+n \in 2 \bZ ,\ \  m\ne n \\
  - \sqrt{3} \, (B_n A_m \mp A_n B_m) &   \quad
  m+n \in 2\bZ + 1
 \end{cases}, \\
\cN^{r, \pm (r-1)}_{nm} & =  \frac{1}{6 (n \mp m)}
 \begin{cases}
  (-1)^{n} (B_n A_m \mp A_n B_m)  & \quad
  m+n \in 2 \bZ ,\ \  m\ne n\\
  \sqrt{3}\, (B_n A_m \pm A_n B_m) &  \quad
  m+n \in 2\bZ +1
 \end{cases}\ .  \nonumber
\end{align}
Observe that the ghost formulae (\ref{eq:ghN}) are related to
matter ones (\ref{eq:matN}) by $A_m \rightarrow - B_m$, $B_m
\rightarrow A_m$.  The ghost Neumann coefficients are expressed via
$\cN^{rs}_{nm}$ as
\begin{subequations}
\begin{align}
 &X^{rs}_{nm} = m \, (\cN^{r,s}_{nm} + \cN^{r, -s}_{nm}) &
 m&\ne n,\, m > 0\, ,
 \label{eq:xxx} \\
 &X^{rr}_{nn}= -\frac{2}{3} (-1)^n A_n B_n + \frac{1}{3}\Bigl(2 \sum_{k=0}^n (-1)^{n-k}A_k^2
  -(-1)^n - A_n^2 \Bigr) & n &\neq 0, \\
 &X^{rs}_{nn}= X^{rs}_{nn} = -\half \left( (-1)^n + X^{rr}_{nn} \right ), &
 r &\neq s , n \neq 0, 
\label{eq:xxx0}
\end{align}
\end{subequations}
The exponential in the vertex $\bra{V_3}$ does not contain $X_{n0}$,
so we have not included an expression for this coefficient;
alternatively, we can simply define this coefficient to vanish and
include $c_0$ in the exponential in $\bra{V_3}$.

Now we describe some algebraic properties satisfied by $V^{rs}$ and
$X^{rs}$.  Define $M^{rs}_{mn} = C V^{rs}$, $\cM^{rs}_{mn} =
\sqrt{\frac{n}{m}} C X^{rs}_{mn}$.  The matrices $M$ and $\cM$
satisfy symmetry and  cyclicity properties
\begin{subequations}
\begin{align}
M^{r+1\, s+1} &= M^{rs},  &  \cM^{r+1\, s+1} &= \cM^{rs},
\label{eq:nccycle} \\
(M^{r s})^T &= M^{r s},  &  (\cM^{r s})^T &= \cM^{rs},
\label{eq:nctranspose} \\
C M^{r s} C &= M^{s r},  &  C \cM^{r s} C &= \cM^{sr}.
\label{eq:ncConjugate}
\end{align}
\end{subequations}
This reduces the set of independent matter Neumann matrices to
$M^{11}$, $M^{12}$, $M^{21}$ and similarly for ghosts.  These
matrices commute and in addition  satisfy
\begin{subequations}
\begin{align}
M^{11}+M^{12}+M^{21} &= - 1, &          \cM^{11}+\cM^{12}+\cM^{21}
&= - 1,
\label{eq:nclinear} \\
M^{12} M^{21} = M^{11}(M^{11}& + 1), &  \cM^{12} \cM^{21} =
\cM^{11}(\cM^{11}&-1).  & \label{eq:ncquadratic}
\end{align}
\end{subequations}
These relations imply that there is only one independent Neumann
matrix.


\begin{thebibliography}{10}

\bibitem{Sen-universality}
A.~Sen,``Universality of the tachyon potential'', JHEP {\bf 9912},
027 (1999), {\tt hep-th/9911116}.

\bibitem{Ohmori:2001am}
K.~Ohmori, ``A review on tachyon condensation in open string field
theories,'' {\tt hep-th/0102085}.

\bibitem{DeSmet:2001af}
P.~J.~De Smet, ``Tachyon condensation: Calculations in string
field theory,'' {\tt hep-th/0109182}.

\bibitem{Zwiebach:nj}
B.~Zwiebach, ``Is The String Field Big Enough?,'' Fortsch.\ Phys.\
{\bf 49}, 387 (2001).

\bibitem{Taylor:2002uv}
W.~Taylor, ``Lectures on D-branes, tachyon condensation, and
string field theory,'' {\tt hep-th/0301094}.

\bibitem{Douglas-Nekrasov}
M.~R.~Douglas and N.~A.~Nekrasov,``Noncommutative field theory'',
Rev.\ Mod.\ Phys.\  {\bf 73}, 977 (2001), {\tt hep-th/0106048}.

\bibitem{Fradkin-Tseytlin85II} E.~S.~Fradkin and A.~A.~Tseytlin,
"Non-linear electrodynamics from quantized strings", \PL {\bf
B163} (1985) 123.

\bibitem{acny}
A.~Abouelsaood, C.~G.~Callan, C.~R.~Nappi and S.~A.~Yost, ``Open
Strings In Background Gauge Fields,'' Nucl.\ Phys.\ B {\bf 280},
599 (1987).

\bibitem{Fradkin-Tseytlin85-rus} E.~S.~Fradkin and A.~A.~Tseytlin,
"Fields as excitations of quantized coordinates", JETP Lett.  41
(1985) 206 (Pisma Z.  Eksp.  Teor.  Fiz.  41 (1985) 169).

\bibitem{Fradkin-Tseytlin85I} E.~S.~Fradkin and ~A.~A.~Tseytlin,
"Effective field theory from quantized strings", \PL {\bf B158}
(1985) 316 , "Quantum string theory effective action", \NP {\bf
B261} (1985) 1


\bibitem{k-Sevrin}
P.~Koerber and A.~Sevrin,
``The non-abelian D-brane effective action through order alpha'**4,''
JHEP {\bf 0210}, 046 (2002)
{\tt hep-th/0208044}.


\bibitem{Seiberg-Witten}  N.\  Seiberg and  E.\  Witten,
``String Theory and Noncommutative Geometry'', JHEP 9909 (1999)
032,{\tt hep-th/9908142}.

\bibitem{Cornalba-Schiappa}
L.~Cornalba and R.~Schiappa,
``Matrix theory star products from the Born-Infeld action,''
Adv.\ Theor.\ Math.\ Phys.\  {\bf 4}, 249 (2000)
{\tt hep-th/9907211}.


\bibitem{Ghoshal-Sen}
D.~Ghoshal and A.~Sen,
``Gauge and general coordinate invariance in nonpolynomial closed string theory,''
Nucl.\ Phys.\ B {\bf 380}, 103 (1992)
{\tt hep-th/9110038}.

\bibitem{David-U}
J.\  R.\ David, ``U(1) gauge invariance from open string field
theory'', {\em JHEP} {\bf 0010} (2000) 017, {\tt  hep-th/0005085}

\bibitem{Sen-action}
A.~Sen,
``Supersymmetric world-volume action for non-BPS D-branes,''
JHEP {\bf 9910}, 008 (1999)
{\tt hep-th/9909062}.


\bibitem{Garousi}
M.~R.~Garousi,
``Tachyon couplings on non-BPS D-branes and Dirac-Born-Infeld action,''
Nucl.\ Phys.\ B {\bf 584}, 284 (2000)
{\tt hep-th/0003122}.

\bibitem{bddep}
E.~A.~Bergshoeff, M.~de Roo, T.~C.~de Wit, E.~Eyras and S.~Panda,
``T-duality and actions for non-BPS D-branes,''
JHEP {\bf 0005}, 009 (2000)
{\tt hep-th/0003221}.


\bibitem{Kluson}
J.~Kluson,
``Proposal for non-BPS D-brane action,''
Phys.\ Rev.\ D {\bf 62}, 126003 (2000)
{\tt hep-th/0004106}.




\bibitem{Ellwood-Shelton-Taylor}
I.~Ellwood, J.~Shelton and W.~Taylor, ``Tadpoles and closed string
backgrounds in open string field theory'', {\tt hep-th/0304259}.

\bibitem{Taylor-amplitudes}
W.~Taylor, ``Perturbative Diagrams in string field
theory", {\tt hep-th/0207132}.

\bibitem{Witten-SFT}
E.\ Witten, ``Non-commutative geometry and string field theory'',
\NP {\bf B268} (1986) 253.

\bibitem{lpp}
A.\ Leclair, M.\ E.\ Peskin and C.\ R.\ Preitschopf, ``String
field theory on the conformal plane (I)'' \NP {\bf B317} (1989) 411-463.

\bibitem{Gaberdiel-Zwiebach}
M.\ R.\ Gaberdiel and B.\ Zwiebach, ``Tensor constructions of open
string theories 1., 2.,'' \NP {\bf B505} (1997) 569, {\tt hep-th/9705038}; \PL {\bf
  B410} (1997) 151, {\tt hep-th/9707051}.

\bibitem{Thorn:1988hm}
C.~B.~Thorn, ``String Field Theory,'' Phys.\ Rept.\  {\bf 175}, 1
(1989).

\bibitem{Gross-Jevicki12}
D.\ J.\ Gross and A.\ Jevicki, ``Operator formulation of
interacting string field theory (I), (II)'', \NP {\bf B283} (1987)
1; \NP {\bf B287} (1987) 225.

\bibitem{cst}
E.\ Cremmer, A.\ Schwimmer and C.\ Thorn,  ''The vertex function
in Witten's formulation of string field theory'', \PL {\bf B179}
(1986) 57.

\bibitem{Samuel}
S.\ Samuel, ''The physical and ghost vertices in Witten's string
field theory'', \PL {\bf B181} (1986) 255.

\bibitem{Ohta}
N.~Ohta, ``Covariant interacting string field theory in the Fock
space representation'', Phys.\ Rev.\ D {\bf 34}, 3785 (1986),
Phys.\ Rev.\ D {\bf 35} (1987), 2627  (E).

\bibitem{Bars}
I.~Bars, I.~Kishimoto and Y.~Matsuo, ``String amplitudes from
Moyal string field theory,''
Phys.\ Rev.\ D {\bf 67}, 066002 (2003), {\tt hep-th/0211131};
``Fermionic ghosts in Moyal string field theory,''
{\tt hep-th/0304005}.



\bibitem{Giddings:1986bp}
S.~B.~Giddings and E.~J.~Martinec, ``Conformal Geometry And String
Field Theory,'' Nucl.\ Phys.\ B {\bf 278}, 91 (1986).

\bibitem{Giddings:1986wp}
S.~B.~Giddings, E.~J.~Martinec and E.~Witten, ``Modular Invariance
In String Field Theory,'' Phys.\ Lett.\ B {\bf 176}, 362 (1986).

\bibitem{Thorn-perttheory}
C.\ B.\ Thorn, ``Perturbation theory for quantized strings'',
\NP {bf B287} (1987) 51-92

\bibitem{Zwiebach:1990az}
B.~Zwiebach, ``A Proof That Witten's Open String Theory Gives A
Single Cover Of Moduli Space,'' Commun.\ Math.\ Phys.\  {\bf 142},
193 (1991).

\bibitem{Kostelecky-Potting}
V.~A.~Kostelecky and R.~Potting, ``Analytical construction
of a nonperturbative vacuum for the open bosonic string'', \PR
{\bf D63:046007} (2001),  {\tt  hep-th/000852}.

\bibitem{Born-Infeld} M.  Born, Proc.  Roy.  Soc.  A 143 (1934) 410;
M.  Born and L.  Infeld, "Foundations of the new field theory",
Proc.  Roy.  Soc.  A 144 (1934) 425; M.  Born, "Th\'{e}orie
non-lin\'{e}are du champ \'{e}lectromagn\'{e}tique", Ann.  inst.
Poincar\'{e}, 7 (1939) 155

\bibitem{Tseytlin}
A.~A.~Tseytlin,
``Born-Infeld action, supersymmetry and string theory,''
{\tt hep-th/9908105}.

\bibitem{Henneaux-db}
G.~Barnich and M.~Henneaux, ``Renormalization Of Gauge Invariant
Operators And Anomalies In Yang-Mills Theory,'' Phys.\ Rev.\
Lett.\  {\bf 72}, 1588 (1994), {\tt hep-th/9312206}; G.~Barnich,
F.~Brandt and M.~Henneaux, ``Local BRST Cohomology In The
Antifield Formalism.  I.  General Theorems.'' Commun.\ Math.\ Phys.\
{\bf 174}, 57 (1995), {\tt hep-th/9405109}; G.~Barnich, F.~Brandt and
M.~Henneaux, ``Local BRST Cohomology In The Antifield Formalism.
II.  Application To Yang-Mills Theory.'' Commun.\ Math.\ Phys.\
{\bf 174}, 93 (1995) {\tt hep-th/9405194};

\bibitem{Henneaux-review}
G.~Barnich, F.~Brandt and M.~Henneaux, ``Local BRST cohomology in
gauge theories,'' Phys.\ Rept.\  {\bf 338}, 439 (2000),
{\tt hep-th/0002245}.

\bibitem{Tseytlin97} A.~A.~Tseytlin,
"On nonabelian generalization of Born-Infeld action in string
theory.",  \NP {\bf B501}:41-52,1997, {\tt hep-th/9701125}.

\bibitem{Hashimoto-Taylor} A.~Hashimoto and W.~Taylor,
"Fluctuation spectra of Tilted and Intersecting D-branes from the
Born-Infeld Action", Nucl.  Phys.  B503 (1997) 193, {\tt hep-th/9703217}.

\bibitem{Bain}
P.~Bain,
``On the non-Abelian Born-Infeld action,''
{\tt hep-th/9909154}.

\bibitem{dst}
F.~Denef, A.~Sevrin and J.~Troost,
``Non-Abelian Born-Infeld versus string theory,''
Nucl.\ Phys.\ B {\bf 581}, 135 (2000)
{\tt hep-th/0002180}.

\bibitem{bsttv}
J.~Bogaerts, A.~Sevrin, J.~Troost, W.~Troost and S.~van der Loo,
``D-branes and constant electromagnetic backgrounds,''
Fortsch.\ Phys.\  {\bf 49}, 641 (2001)
{\tt hep-th/0101018}.


\bibitem{Neveu-Scherk}
A.\ Neveu and J.\ Scherk, ``Connection between Yang-Mills fields and
dual models'', 
\NP {\bf B36} (1972) 155-161;

\bibitem{Scherk-Schwarz}
 J.\ Scherk and J.H.\ Schwarz, 
``Dual models for non-hadrons'', \NP {\bf B81} (1974) 118-144.

\bibitem{Tseytlin86} A.~A.~Tseytlin,
"Vector field effective action in the open superstring theory",
\NP {\bf B276} (1986) 391; \NP{\bf B291} (1987) 876 (errata);


\bibitem{Polchinski} J.~Polchinski, ``String Theory '', Cambridge
University Press, Cambridge, England, 1998.



\bibitem{Kostelecky-Samuel}
V.~A.~Kostelecky and S~.~Samuel, ``On a nonperturbative
vacuum for the open bosonic string'', \NP {\bf B336} (1990)
263-296.

\bibitem{Sen-Zwiebach}
A.~Sen and B.~Zwiebach, ``Tachyon condensation in string field
theory,'' JHEP {\bf 0003}, 002 (2000), {\tt hep-th/9912249}.

\bibitem{Moeller-Taylor} N.~Moeller and W.~Taylor,
``Level truncation and the tachyon in open bosonic string field
theory'', \NP {\bf B583} (2000) 105-144,  {\tt hep-th/0002237}.

\bibitem{Berkovits-Schnabl} N.\ Berkovits and M.\ Schnabl, to appear.

\bibitem{Taylor-action}
W.\ Taylor, ``D-brane effective field theory from string field
theory'', \NP  {\bf B585} (2000) 171-192

\bibitem{Grigoriev-Henneaux}
G.~Barnich, M.~Grigoriev and M.~Henneaux ``Seiberg-Witten
maps from the point of view of consistent  deformations of gauge
theories'', ULB-TH/01-17, {\tt hep-th/0106188}.

\bibitem{Henneaux-Teitelboim} M.~Henneaux and C.~Teitelboim
``Quantization of Gauge Systems'', Princeton University Press,
1992.

\bibitem{equivalencetheorem}
A.\ Salam and J.\ Strathdee, \PR {\bf D2} (1970) 2869, R.  \
Kallosh and I.\ V.\ Tyutin, {\em Yad.  Fiz}.  {\bf 17} (1973) 190,
Y.\ M.\ P.\ Lam \PR {\bf D7} (1973) 2943, M.\ C.\ Berger and
Y.\ M.\ P.\ Lam, \PR  {\bf D13} (1975) 3247

\bibitem{Liu}
H.~Liu,
``*-Trek II: *n operations, open Wilson lines and the Seiberg-Witten  map,''
Nucl.\ Phys.\ B {\bf 614}, 305 (2001)
{\tt hep-th/0011125}.

\bibitem{van-Raamsdonk}
M.~Van~Raamsdonk,
``Blending local symmetries with matrix nonlocality in D-brane effective  actions,''
{\tt hep-th/0305145}.


\bibitem{Hashimoto-Itzhaki}
A.~Hashimoto and N.~Itzhaki,
``Observables of string field theory,''
JHEP {\bf 0201}, 028 (2002)
{\tt hep-th/0111092}.

\bibitem{Gaiotto-Rastelli-Sen-Zwiebach}
D.~Gaiotto, L.~Rastelli, A.~Sen and B.~Zwiebach,
``Ghost structure and closed strings in vacuum string field theory,''
{\tt hep-th/0111129}.

\bibitem{Berkovits}
N.~Berkovits,
``SuperPoincare invariant superstring field theory,''
Nucl.\ Phys.\ B {\bf 450}, 90 (1995)
[Erratum-ibid.\ B {\bf 459}, 439 (1996)]
{\tt hep-th/9503099}.

\bibitem{Witten:1986qs}
E.~Witten,
``Interacting Field Theory Of Open Superstrings,''
Nucl.\ Phys.\ B {\bf 276}, 291 (1986).

\bibitem{Preitschopf:1989gp}
C.~R.~Preitschopf, C.~B.~Thorn and S.~A.~Yost,
``Superstring Field Theory,''
UFIFT-HEP-90-3
{\it Invited talk given at Workshop on Superstring and Particle
Theory, Tuscaloosa, AL, Nov 8-11, 1989}

\bibitem{Arefeva:1989cm}
I.~Y.~Arefeva, P.~B.~Medvedev and A.~P.~Zubarev,
Phys.\ Lett.\ B {\bf 240}, 356 (1990).


\bibitem{Ohmori}
K.~Ohmori,
``Level-expansion analysis in NS superstring field theory revisited,''
{\tt hep-th/0305103}.

\bibitem{Kutasov-Niarchos}
D.~Kutasov and V.~Niarchos,
``Tachyon effective actions in open string theory,''
{\tt hep-th/0304045}.


\bibitem{llm}
N.~Lambert, H.~Liu and J.~Maldacena,
``Closed strings from decaying D-branes,''
{\tt hep-th/0303139}.

\end{thebibliography}
\end{document}